\documentclass[english,11pt,aps,prd,eqsecnum]{revtex4}
\usepackage{amssymb,amsmath,amsthm,graphicx,amscd}
\usepackage{enumerate,color,comment,ulem,bm,floatrow,wrapfig,caption,subcaption}
\usepackage{xcolor}
\usepackage[papersize={8.5in,11in}]{geometry}
\usepackage[hidelinks]{hyperref}

\usepackage{orcidlink}

\begin{document}

\title{Atom-Field-Medium Interactions II:  Covariance Matrix Dynamics for \\ $N$ Harmonic Atoms in a Dielectric-Altered Quantum Field and \\Effects of Dielectric on Atom-Field Entanglement}
\author{Jen-Tsung Hsiang\orcidlink{0000-0002-9801-208X}}
\email{cosmology@gmail.com}
\affiliation{College of Electrical Engineering and Computer Science, National Taiwan University of Science and Technology, Taipei City, Taiwan 106, R.O.C.}
\author{Bei-Lok Hu\orcidlink{0000-0003-2489-9914}}
\email{blhu@umd.edu}
\affiliation{Maryland Center for Fundamental Physics and Joint Quantum Institute,  University of Maryland, College Park, Maryland 20742, USA}

\date{March 17, 2025}

\begin{abstract}
We continue our investigation of multi-partite open quantum systems comprising layers of structure using the atom-field-medium interactions as a familiarly important example. Same as in  Paper I \cite{HH24}  we consider a system of $N$ harmonic oscillators, modeling the internal degrees of freedom (idf) of $N$ neutral atoms interacting with a scalar quantum field altered by the presence of a dielectric medium. Different from Paper I, which uses the graded influence action formalism, here, taking advantage of the Gaussian nature of our extended system's interactions, we use the quantum Langevin equation method to calculate the time evolution of the covariance matrix elements of the quantum correlation functions of the idfs of the $N$ system-atoms in a dielectric-altered quantum field. The covariance matrix is particularly useful for extracting quantum informational properties of a Gaussian system related to quantum correlations, such as quantum entanglement. As an illustration of the method we calculate the entanglement between one system atom and the ambient quantum field outside the dielectric half-space,  measured  by the purity function and the von Neumann entropy. We highlight one somewhat peculiar feature in our results and one important technical issue:  The special feature refers to the non-monotonic behavior of the purity function when the atom is positioned very close to the dielectric surface. By deriving the Robertson-Schr\"odinger function and displaying a similar qualitative behavior under these conditions we attribute this novelty to a manifestation of the uncertainty relation. The technical issue refers to the order-reduction scheme to remove the third time derivative term in the Langevin equation for the idfs of the atom.  We point out the inconsistencies in the traditional treatments and propose  a new consistent scheme of order reduction for Gaussian open systems.  
\end{abstract}

\maketitle

\hypersetup{linktoc=all}
\baselineskip=18pt
\allowdisplaybreaks

\tableofcontents

\section{Introduction}
In a recent paper \cite{HH24} (Paper I) we developed the graded influence action formalism introduced earlier by Behunin and Hu \cite{BH10,BH11} for treating multi-partite open quantum systems comprising several layers of structure with self-consistent back-actions and applied it to the study of atom-field-medium interactions. We considered a system of $N$ harmonic oscillators, modeling the internal degrees of freedom (idf) of $N$ neutral atoms (A), interacting with a scalar quantum field (F),  altered by the presence of a dielectric medium (M),  and showed how to derive the stochastic equations for  the dynamical variables  {of the reduced subsystems} in the successive layers of structure. The end result obtained there is an equation of motion which describes the nonequilibrium stochastic dynamics of the idf of the atoms interacting with a dielectric medium-modified quantum field. By studying the internal dynamics of these neutral atoms, as probes or as measuring devices, one can extract the quantum correlations or   the statistical properties of the ambient field, study the  ramifications or extensions of the Casimir-Polder effect~\cite{Ca48,CP48,Pe97,Ba10a,Ba10b,Bu12,Bu13,BH09,BH10,BH11}, or uncover the hidden information  in the underlying lower-tier structures.  In this paper, within the same conceptual framework and structural scheme of the graded actions, we present a different route, via quantum Langevin equations \cite{FLO88,FK87,HH19c}, and calculate the time evolution of the covariance matrix elements \cite{NC11,Ad07} of the quantum correlation functions of the idf of $N$-atoms in a dielectric-altered quantum field. 

%\subsubsection{Functional formalism versus Quantum Langevin Equation methods}

The quantum Langevin equations (QLE), derived from a simultaneous set of Heisenberg operator equations of subsystems that constitute a multi-partite interacting linear systems, offers an equivalent description to the influence functional (IF) formalism, presented in Paper I. Comparing the two approaches applied to Gaussian systems, the quantum Langevin equation method has the advantages that it is more intuitive, and physically transparent to describe the dynamics of the reduced system, due to the resemblance to its classical counterpart. It is easier to see how the influences of the underlying subsystems of a lower tier stack up in affecting the higher tier subsystems.   Once we have solved the Langevin equation, it is much easier to compute the physical quantities of the reduced system than using the influence functional formalism, and the initial state of the reduced system is not restricted to be Gaussian. In terms of disadvantage, the QLE approach has the shortcoming of treating nonlinear systems due to noncommutativity of operators and potential ambiguities arising from operator ordering.

To enable obtaining analytical solutions as much as possible, we make two reasonable assumptions in computing the entanglement measure in Sec.~V. First, we examine only the late-time quantum correlations among the idf of the $N$ atoms  located outside the dielectric medium.  The late-time regime is needed to allow the dielectric-modified quantum field to reach an equilibrium,  so that the $N$ atoms will interact with an equilibrated modified field~\footnote{Dynamical equilibration for such an interacting multipartite systems, though likely for a wide class of Guassian systems, is not guaranteed (e.g., see discussions in~\cite{HH18a} for a bi-partite system and~\cite{Ve11}.), and its justification requires a full investigation worthy of another paper. We shall not take this for granted and thus call it an assumption.}. Second, we will not consider the contributions from the bounded polarizable source $\varrho$ inside the dielectric, that is, the field fluctuations generated by the intrinsic dipoles of the dielectric atoms and the dipoles of the dielectric atoms induced by its thermal bath.  Other than the linearity condition for all interactions at each level and these two assumptions, the formal results derived here are rather general, with little reference to the details of the ingredients. 

The covariance matrix is particularly useful for extracting quantum informational properties of the  {Gaussian} system related to quantum correlations, such as quantum entanglement~\cite{LH09,HH15b,Rong,HL06,HF04,DR22,HH21,AS04,AS05,AI05,PI03,Pl05,Si00,AR14,Ad07,Pe96,HH96,AH23,HA22}. As an illustration of the method we apply our results to calculate the ``purity"~\cite{NC11,AS05,PI03} between one such system atom and a dielectric half-space, mediated by an ambient quantum field.

For a system comprised of an arbitrary number of quantum harmonic oscillators linearly coupled to a common bath, here represented by a quantum scalar field as the system's environment,  with a  quadratic action,  this problem can be solved exactly. (See, e.g, \cite{HPZ92} for one quantum harmonic oscillator motion in a $N$-quantum harmonic oscillator bath. We have collected the essentials in the analysis of the  motion of a multi-partite quantum open system in Appendix A.)    {Following this line of thought,} we derive the wave equation  of the dielectric-modified quantum field, in which we identify the dissipation kernel associated with the dielectric atoms as the susceptibility of the dielectric medium, after an appropriate averaging process has been introduced. From this we  {can further} obtain the Langevin equation governing the reduced system of the idf of the $N$ atoms. Under the {assumed} equilibrium conditions for {the internal dynamics of the system atoms}, such as likely be the case at late times,  one can  derive an exact form for the time-dependent (nonequilibrium) covariance matrix elements solely based on the dissipation kernels, which are the retarded Green's functions of the internal dynamics of the atoms.  

In Sec. II we shall present the details in our derivation of the Langevin equation for the idf of the $N$-atom open quantum system. {The material is pretty self-contained. Noteworthy is the way an order reduction scheme is applied to the equations of motion of a reduced system. We point out the inconsistencies in the traditional treatments and present  a new consistent scheme of order reduction for Gaussian open systems.  Details of our prescription is contained in Appendix B. } In Sec. III we give a short summary of the connection between covariance matrix and entanglement measures. In Sec. IV we   present  the Green functions of the quantum field modified by a dielectric half space. In Sec. V we assume that at late times the field has settled in equilibrium and derive the covariance matrix dynamics of the idf of the system atoms interacting with a dielectric-modified field.  We then apply our results to the calculation of the purity function as a measure of the entanglement between one such system atom and the dielectric half space. We show the entanglement behavior for different parameters, amongst which of special interest is a non-monotonic feature when the atom is located very close to the dielectric surface. We offer corroborative evidence for such a phenomenon by displaying similar qualitative behavior in the Robertson-Schr\"odinger relation. {In Sec. VI we end with a summary of our key findings and points of interest worthy of  further investigations.} 
%in the derivation of the dynamical equations of motion and in the qualitative behaviors of atom-dielectric entanglement.

\section{Quantum Langevin Equation Treatment of Atom-Field-Dielectric Interaction}

In Paper I our emphasis was placed on the derivation and analysis of the structures of the graded influence actions, showing how noises in a  Gaussian subsystem at a lower tier (e.g., the medium with its bath,  modifying the ambient quantum field ) enter in the action and the equations of motion of a subsystem at a higher tier (e.g., the internal degrees of freedom such as the electric dipoles of the $N$ neutral atoms). We show at the end how to calculate the more directly measurable quantities like the covariant matrix, and from there, various useful physical quantities such as purity as a measure of atom-dielectric entanglement.

In this paper we shall expound the route via quantum Langevin equations, which usually is easier than the path integral method and perhaps more transparent in attributing the physical meanings because for Gaussian systems the equations of motion for the quantum operators have the same forms as their classical counterparts. The drawback is the difficulty in treating nonlinear quantum systems, whereas for functional integral methods, perturbative schemes can be more easily applied to treating systems with weak nonlinearity~\cite{HPZ93,HH20a,HH20b,YH20}.

\subsection{Oscillator model for absorptive dielectric medium} 

As was shown in Paper I and its precursors \cite{BH10,BH11} and references mentioned therein, to account for absorption in the medium we need to introduce a common bath coupled to all atoms of the dielectric medium. We model it by a massless quantum scalar field $\psi$, which  only lives in the medium, and its dynamics is described by a free-field action 
\begin{equation}
	S_{ \textsc{b}}[\psi]=\frac{1}{2}\int\!d^{4}x\; (\partial_{\mu}\psi)\,(\partial^{\mu}\psi)\,,
\end{equation}
plus an interaction action of it with the dielectric atom with coupling strength $g_{ \textsc{b}}$:   
\begin{equation}
	S_{\textsc{mr}}[\varsigma,\phi]=g_{ \textsc{b}}\int\!dt\,d^{3}\bm{x}\;\sum_{i}\delta^{3}(\bm{x}-\bm{a}_{i})\,\varsigma(\bm{a}_i,t)\,\psi(\bm{x},t)\,,
\end{equation}
where $\varsigma(\bm{a}_i,t)$ is the dynamic variable representing the internal degrees of freedom of the $i^{\text{th}}$ neutral dielectric atom located at $\bm{a}_i$.

The internal degrees of freedom $\varsigma(\bm{a}_i,t)$ of a neutral dielectric atom is modeled by a quantum harmonic oscillator whose dynamics describes the time-dependent electronic configuration of the atom. Its action has the form
\begin{equation}
    S_{\textsc{m}}[\varsigma]=\frac{m}{2}\int\!dt\,d^{3}\bm{x}\sum_{i}\delta^{3}(\bm{x}-\bm{a}_{i})\,\Bigl[\dot{\varsigma}^{2}(\bm{a}_i,t)-\varpi^{2}\varsigma^{2}(\bm{a}_i,t)\Bigr]\,.
\end{equation}
where $m$ is the (bare) mass and $\varpi$ is the (bare) natural frequency of the oscillator.

The dielectric medium interacts with our system of interest, that made up of $N$ harmonic atoms, by way of an ambient massless scalar field $\phi$, which in our model plays the role of a specified polarization of an electromagnetic field. Note that it is distinct from the bath field in both its setting and its function. The action of the ambient field and its coupling with the dielectric are given by 
\begin{align}
    S_{\textsc{f}}[\phi]&=\frac{1}{2}\int\!d^{4}x\;( \partial_{\mu}\phi)\,(\partial^{\mu}\phi)\,,\\
    S_{\textsc{mf}}[\varsigma,\phi]&=g_{\textsc{m}}\int\!dt\,d^{3}\bm{x}\;\sum_{i}\delta^{3}(\bm{x}-\bm{a}_{i})\,\varsigma(\bm{a}_i,t)\dot{\phi}(\bm{x},t)\,,
\end{align}
with the coupling strength $g_{\textsc{m}}$. This type of coupling in the interaction action $S_{\textsc{mf}}$, called a ``derivative coupling",  is chosen to mimic the electric dipole interaction $-\bm{p}\cdot\bm{E}$ with the electromagnetic field.

Finally, we come to our system of $N$ neutral atoms, where the $n^{\text{th}}$ atom is located at $\bm{z}_n$, in the space outside the dielectric medium. Their   internal degrees of freedom modeled by harmonic oscillators of mass $M$ and natural frequency $\Omega$ described by the action
\begin{equation}
    S_{\textsc{a}}[\chi^{(n)}]=\frac{M}{2}\int\!dt\;\Bigl[\dot{\chi}^{(n)2}-\Omega^{2}\,\chi^{(n)2}\Bigr]\,,
\end{equation}
Their interaction with the medium-modified ambient field with coupling strength $g_{\textsc{f}}$ is accounted for by the action
\begin{equation}
     S_{\textsc{af}}[\{\chi\},\phi]=g_{\textsc{f}}\sum_{n=1}^{N}\int\!dt\;\chi^{(n)}(t)\,\dot{\phi}(\bm{z}_{n},t)\,,
\end{equation}
where $\chi^{(n)}$, $\bm{z}_n$ are the internal (electronic) and the external (mechanical) degrees of freedom of the $n^{\text{th}}$ atom.

We can readily write down a simultaneous set of Heisenberg (operator) equations of motion for these four interacting subsystems
\begin{align}
    \ddot{\hat{\psi}}(\bm{x},t)-\bm{\nabla}^2\hat{\psi}(\bm{x},t)&=g_{ \textsc{b}}\sum_{i}\delta^{3}(\bm{x}-\bm{a}_{i})\,\hat{\varsigma}(\bm{a}_i,t)\,,\label{E:nbks1}\\
    m\,\ddot{\hat{\varsigma}}(\bm{a}_i,t)+m\varpi^2\hat{\varsigma}(\bm{a}_i,t)&=g_{ \textsc{b}}\hat{\psi}(\bm{a}_i,t)+g_{\textsc{m}}\dot{\hat{\phi}}(\bm{a}_i,t)\,,\label{E:nbks2}\\
    \ddot{\hat{\phi}}(\bm{x},t)-\bm{\nabla}^2\hat{\phi}(\bm{x},t)&=-g_{\textsc{m}}\sum_{j}\delta^{3}(\bm{x}-\bm{a}_{j})\,\dot{\hat{\varsigma}}(\bm{a}_j,t)-g_{\textsc{f}}\sum_{n}\delta^{3}(\bm{x}-\bm{z}_{n})\,\dot{\hat{\chi}}^{(n)}(\bm{z}_n,t)\,,\label{E:nbks3}\\
    M\,\ddot{\hat{\chi}}^{(n)}(t)+M\Omega^2\,\hat{\chi}^{(n)}(t)&=g_{\textsc{f}}\sum_n\dot{\hat{\phi}}(\bm{z}_n,t)\,.\label{E:nbks4}
\end{align}
We use the same notations as in Paper I %Sec.~\ref{S:beei} of Paper I.

\subsection{Dynamics of dielectric atoms}

Solving Eq.~\eqref{E:nbks1} gives
\begin{equation}\label{E:nvkije}
    \hat{\psi}(\bm{x},t)=\hat{\psi}_h(\bm{x},t)+g_{ \textsc{b}}\int^t_0\!\!dt'\!\int\!d^3\bm{y}\;G_{\mathrm{R},0}^{(\psi_h)}(\bm{x},t;\bm{y},t')\,\sum_i\delta^{(3)}(\bm{y}-\bm{a}_{i})\,\hat{\varsigma}(\bm{a}_i,t')\,,
\end{equation}
where $\hat{\psi}_h(\bm{x},t)$, satisfying the source-free field equation, is the homogeneous solution to the wave equation~\eqref{E:nbks1}. %This corresponds to $\xi_0^{(\psi_h)}$ in Eq.~\eqref{E:vdk}. 
The second term on the right gives the inhomogeneous solution and can be viewed as the radiation field emitted from each dielectric atom back to the bath. Their physical meaning will be better seen later. In the nonlocal expression, the kernel function $G_{\mathrm{R},0}^{(\psi_h)}(\bm{x},t;\bm{y},t')$ is the retarded Green's function associated with the free bath field $\hat{\psi}_h(\bm{x},t)$ which carries the causal influences between the dielectric atoms, mediated by the bath field. Formally, we could interpret $\hat{G}^{(\psi_h)}_{\mathrm{R},0}$ as $\square^{-1}$ with $\square=\partial_t^2-\bm{\nabla}^2$ in the operator sense.

Substituting \eqref{E:nvkije} back to Eq.~\eqref{E:nbks2}, we obtain the equation of motion for the dielectric atoms idf $\varsigma$,
\begin{align}\label{E:bktrr}
     m\ddot{\hat{\varsigma}}(\bm{a}_i,t)+m\varpi^2\hat{\varsigma}(\bm{a}_i,t)&=g_{\textsc{m}}\dot{\hat{\phi}}(\bm{a}_i,t)+g_{ \textsc{b}}\hat{\psi}_h(\bm{a}_i,t)+g_{ \textsc{b}}^2\sum_j\int^t_0\!dt'\;G_{\mathrm{R},0}^{(\psi_h)}(\bm{a}_i,t;\bm{a}_j,t')\,\hat{\varsigma}(\bm{a}_j,t')\,.
\end{align}
Written in this way, the bath variable has disappeared but its effects are fully accounted for by $\hat{\psi}_h(\bm{x},t)$ and the integral expression on the right hand side. It would be instructive to decompose its contribution into two parts  $j=i$ or otherwise. The $j=i$ case gives the radiation reaction and the renormalization of the oscillating frequency $\varpi$ from the bare quantity  to the physical frequency   $\varpi_{\varsigma}$ of $\varsigma$. The former will lead to a damping term on the left hand side of Eq.~\eqref{E:bktrr}. On the other hand, when $j\neq i$, as seen from the expression and the retardation nature of $G_{\mathrm{R},0}^{(\psi_h)}$, it describes the causal, spatially non-Markovian influences between dielectric atoms. Following this interpretation, we may re-write \eqref{E:bktrr} in a physically more transparent expression
\begin{align}\label{E:bqwrr}
     \ddot{\hat{\varsigma}}(\bm{a}_i,t)+2\gamma_{\varsigma}\,\dot{\hat{\varsigma}}(\bm{a}_i,t)+\varpi^2_{\varsigma}\,\hat{\varsigma}(\bm{a}_i,t)&-\frac{g_{ \textsc{b}}^2}{m}\sum_{j\neq i}\int^t_0\!dt'\;G_{\mathrm{R},0}^{(\psi_h)}(\bm{a}_i,t;\bm{a}_j,t')\,\hat{\varsigma}(\bm{a}_j,t')\notag\\
     &=\frac{g_{\textsc{m}}}{m}\,\dot{\hat{\phi}}(\bm{a}_i,t)+\frac{g_{ \textsc{b}}}{m}\,\hat{\psi}_h(\bm{a}_i,t)\,,
\end{align}
where $\gamma_{\varsigma}=g_{ \textsc{b}}^2/8\pi m$ is the damping constant. Thus the dissipative behavior of the dielectric emerges. On the right hand side, $\hat{\psi}_h(\bm{a}_i,t)$ recounts the free quantum fluctuations from the bath and thus serves as a noise force from the bath on the dielectric atoms at $\bm{a}_i$. There is another forcing term $g_{\textsc{m}}\,\dot{\hat{\phi}}(\bm{a}_i,t)$, which is analogous to the electromagnetic Lorentz force acting on the charge distribution in the dielectric atom. Note that the field $\hat{\phi}(\bm{a}_i,t)$ is an interacting field coming from a complete solution to Eq.~\eqref{E:nbks3}, not a free field like $\hat{\psi}_h(\bm{a}_i,t)$, which is the homogeneous solution to the corresponding wave equation, Eq.~\eqref{E:nbks1}.

This is a good place to comment on the dynamics described by Eq.~\eqref{E:bqwrr} which will be important to interpret the medium-modified field. Generically speaking, the solution to Eq.~\eqref{E:bqwrr} does not have a closed form~\cite{HH18a} due to the nonlocal terms. It can still be decomposed into a homogeneous contribution which solely depends on the initial conditions of $\hat{\varsigma}$, and an inhomogeneous contribution that is driven by the right hand side of Eq.~\eqref{E:bqwrr}. Under suitable conditions~\cite{Ve11,HH18a} concerning the stability of dynamics, the homogeneous contribution decays with time, so that at late times only the inhomogeneous contribution survives. This is most easily seen when the nonlocal terms are negligible. The homogeneous component, due to the damping, decays exponentially at a rate related to $\gamma_{\varsigma}$. The presence of the nonlocal term usually makes damping weaker~\cite{HA22,HH22}, a signature of non-Markovian effects that the lingering system memory can often sustain the coherence in a system for an extended period of time. Since the nonlocal term scales with a certain inverse power of the separation between the dielectric atoms and depends on the relative phase between them, its contribution can be relatively small for a dilute dielectric medium. However, its presence is essential in maintaining the self-consistency of the underlying theory and plays an important role in the system's ability to attain dynamical equilibration~\cite{HH18a,HH19a}.

Explicitly, if the nonlocal term in Eq.~\eqref{E:bqwrr} indeed can be ignored, then the solution of $\hat{\varsigma}$ is greatly simplified because Eq.~\eqref{E:bqwrr} now describes a driven damped harmonic oscillator and accounts for the usual behavior of the charge in the Lorentz model of the dielectric. The solution takes on a generic form
\begin{equation}\label{E:rjkd}
    \hat{\varsigma}(\bm{a}_i,t)=\hat{\varsigma}_h(\bm{a}_i,t)+\int_0^t\!dt'\;G_{\mathrm{R}}^{(\varsigma_h)}(\bm{a}_i,t;\bm{a}_i,t')\,\hat{J}(\bm{a}_i,t')\,,
\end{equation}
where the totality $\hat{J}(\bm{a}_i,t)$ of the fluctuating sources is 
\begin{equation}
    \hat{J}(\bm{a}_i,t)=g_{\textsc{m}}\,\dot{\hat{\phi}}(\bm{a}_i,t)+g_{ \textsc{b}}\,\hat{\psi}_h(\bm{a}_i,t)\,,
\end{equation}
and $G_{\mathrm{R}}^{(\varsigma_h)}(\bm{a}_i,t;\bm{a}_i,t')$ is the retarded Green's function for Eq.~\eqref{E:bqwrr} evaluated at the position $\bm{a}_i$ of the $i^{\text{th}}$ dielectric atom in the absence of the nonlocal term. In this case, it is explicitly given by
\begin{equation}\label{E:hgurte}
    G_{\mathrm{R}}^{(\varsigma_h)}(\bm{a}_i, t;\bm{a},t')=\frac{1}{mw_{\varsigma}}\,e^{-\gamma_{\varsigma}(t-t')}\,\sin w_{\varsigma}(t-t')\,,\qquad\qquad w_{\varsigma}=\sqrt{\smash{\varpi_{\varsigma}^2}-\gamma_{\varsigma}^2}\,.
\end{equation}
The important observation here is that $\hat{\varsigma}_h(\bm{a}_i,t)$ in Eq.~\eqref{E:rjkd} decays like $e^{-\gamma_{\varsigma}t}$. This will recount the intrinsic, spontaneous dipole moments of the neutral atoms which make up the dielectric medium. For time scales much greater than $\gamma_{\varsigma}^{-1}$, they diminish to a negligible extent. What is left is the induced moment enacted by the interaction with the bath and the ambient field, as shown in the second term on the right hand side of Eq.~\eqref{E:rjkd}. Finally, the damping term in Eq.~\eqref{E:rjkd} depicts the absorptive behavior of the dielectric, as in the Lorentz model. %As a reminder, the homogeneous term $\hat{\varsigma}_h$ in Eq.~\eqref{E:rjkd} corresponds to the stochastic source $\xi^{(\varsigma_h)}$ in Eq.~\eqref{E:dmkea}.

In general, the solution to Eq.~\eqref{E:bqwrr} has a more complicated form
\begin{equation}
    \hat{\varsigma}(\bm{a}_i,t)=\hat{\varsigma}_h(\bm{a}_i,t)+\sum_j\int_0^t\!dt'\;G_{\mathrm{R}}^{(\varsigma_h)}(\bm{a}_i,t;\bm{a}_j,t')\,\hat{J}(\bm{a}_j,t')\,,
\end{equation}
in which the nonlocal term includes the contributions originating from the positions of other dielectric atoms. In this case, the formal form of $G_{\mathrm{R}}^{(\varsigma_h)}(\bm{x}_i,t;\bm{a}_j,t')$ is similar to that of $G_{\mathrm{R}}^{(\chi_h)}(\bm{z}_n,t;\bm{z}_n,t')$ later given by the inverse Laplace transform of Eq.~\eqref{E:oiuehbd}.

Hitherto we have argued that in the neutral atoms which constitute the dielectric body, the contribution of the intrinsic dipole moments will gradually vanish with time. The majority contribution in the dipoles comes from the induced components. In the absence of the ambient field, from the viewpoint of nonequilibrium evolution of the interacting system, it can be shown that this driven dynamics of the dipole moments will reach equilibration eventually when the average energy exchange with the bath is balanced.

However, the subtle issue is that this final equilibrium state is not necessarily a Gibbs state, i.e., a thermal state. Only when the coupling between the dielectric atom and the thermal bath is vanishingly small will the equilibrium state be a thermal one. In this special limit, the final density matrix of the dipole moment looks like the standard one for its thermal state. At this stage, although these dipole moments have undergone nonequilibrium evolution, their dynamics  manifests as if they are in thermal equilibrium all along. {Thus, in this special yet often invoked limit,  one cannot quite distinguish  between the results from an equilibrium closed-system and a nonequilibrium open-system treatment.  This is the presumed condition and unspoken assumption behind many rigorous  theories} of  absorptive medium in a thermal environment.

In a more general setting where the coupling between the dielectric and the thermal bath is not diminishing, we cannot assume that the dipole moments will rest in a thermal state. The density matrix will still take a Gaussian form for the linear open-system dynamics under consideration, and their elements can be easily expressed in terms the covariance density matrix elements, introduced in Sec.~\ref{S:eeuidf}.

\subsection{Dynamics of ambient field}

With sufficient knowledge of the dynamics of the internal degrees of freedom of the dielectric atoms, we may proceed to examine the dynamics of the ambient field modified by the absorptive dielectric medium.  So far we have obtained
\begin{align}
    &\hat{\varsigma}(\bm{a}_i,t)=\hat{\varsigma}_h(\bm{a}_i,t)+\sum_j\int_0^t\!dt'\;G_{\mathrm{R}}^{(\varsigma_h)}(\bm{a}_i,t;\bm{a}_j,t')\,\hat{J}(\bm{a}_j,t')\,,\qquad\qquad \hat{J}(\bm{x},t)=g_{\textsc{m}}\,\dot{\hat{\phi}}(\bm{x},t)+g_{ \textsc{b}}\,\hat{\psi}_h(\bm{x},t)\,,\notag\\
    &\partial^2_t\hat{\phi}(\bm{x},t)-\bm{\nabla}^2\phi(\bm{x},t)=-g_{\textsc{m}}\sum_{i}\delta^{3}(\bm{x}-\bm{a}_{i})\,\dot{\hat{\varsigma}}(\bm{a}_i,t)-g_{\textsc{f}}\sum_{n}\delta^{3}(\bm{x}-\bm{z}_{n})\,\dot{\hat{\chi}}^{(n)}(\bm{z}_n,t)\,,\label{E:nbks6}
\end{align}
from Eqs.~\eqref{E:rjkd} and \eqref{E:nbks3}. Combining them together, we find
\begin{align}\label{E:beoitue}
    &\quad\partial^2_t\hat{\phi}(\bm{x},t)-\bm{\nabla}^2\hat{\phi}(\bm{x},t)+g_{\textsc{m}}^2\sum_{i,j}\delta^{3}(\bm{x}-\bm{a}_{i})\,\partial_t\int_0^t\!dt'\;G_{\mathrm{R}}^{(\varsigma_h)}(\bm{a}_i,t;\bm{a}_{j},t')\,\partial_{t'}\hat{\phi}(\bm{a}_{j},t')\notag\\
    &=-g_{\textsc{m}}\sum_{i}\delta^{3}(\bm{x}-\bm{a}_{i})\,\dot{\hat{\varsigma}}_h(\bm{a}_i,t)-g_{\textsc{m}}g_{ \textsc{b}}\sum_{i,j}\delta^{3}(\bm{x}-\bm{a}_{i})\,\partial_t\int_0^t\!dt'\;G_{\mathrm{R}}^{(\varsigma_h)}(\bm{a}_i,t;\bm{a}_{j},t')\,\hat{\psi}_h(\bm{a}_{j},t')\notag\\
    &\qquad\qquad\qquad\qquad-g_{\textsc{f}}\sum_{n}\delta^{3}(\bm{x}-\bm{z}_{n})\,\dot{\hat{\chi}}^{(n)}(\bm{z}_n,t)
\end{align}
The nonlocal term on the left hand side tells how the ambient field at $\bm{x}$ is modified by the electric dipoles of the dielectric atoms at $\bm{a}_j$ induced by the ambient field there. The first term on the right hand is nothing but the intrinsic component of the dipole of the dielectric atom at $\bm{a}_i$ that acts as the source to generate the ambient field. Its contribution, as argued earlier regarding the dynamics of $\hat{\varsigma}$, tends to decay with time due to the damping caused by the interaction with the bath field $\hat{\psi}$. The second term is the contribution from the induced dipole  by the thermal bath $\hat{\psi}_h$. They together will be called the bounded polarizable source, denoted by $\hat{\varrho}(\bm{x},t)$. From the construction, it is easy to see that they are nonzero only ``inside''~\footnote{This assumes all the dielectric atoms are sufficiently bounded and occupy a well defined region.} the dielectric medium. The final term on the right hand side accounts for the dipole source from the $N$ atoms outside the dielectric.

Eq.~\eqref{E:beoitue} is the microscopic counterpart of the  wave equation in a  macroscopic description of the electromagnetic field in the presence of the dielectric material. For the moment, let us ignore the contribution of the $N$ atoms. Then following the preceding discussions, we can easily see how and why the field is modified by the medium. This connection is better seen in Fourier space, where we can cast Eq.~\eqref{E:beoitue} into a familiar form
\begin{align}\label{E:ndeiee}
    \bm{\nabla}^2\tilde{\hat{\phi}}(\bm{x},\omega)+\omega^2\tilde{\hat{\phi}}(\bm{x},\omega)+\int\!d^3\bm{y}\;\omega^2\tilde{\chi}_e(\bm{x},\bm{y};\omega)\,\tilde{\hat{\phi}}(\bm{y},\omega)=-\tilde{\hat{\varrho}}(\bm{x},\omega)\,.
\end{align}
Here the Fourier transform of a function $f(t)$ of $t$ is defined by
\begin{equation}
    \tilde{f}(\omega)=\int_{-\infty}^{\infty}\!dt\;e^{i\omega t}f(t)\,,
\end{equation}
such that
\begin{equation}
    \int_{-\infty}^{\infty}\!dt\;e^{i\omega t}\int_{-\infty}^{\infty}\!ds\;f(t-s)g(s)=\tilde{f}(\omega)\tilde{g}(\omega)\,.
\end{equation}

The susceptibility tensor $\tilde{\chi}_e(\bm{x},\bm{y};\omega)$ 
\begin{equation}\label{E:eudwisd}
    \tilde{\chi}_e(\bm{x},\bm{y};\omega)=g_{\textsc{m}}^2\sum_{j}\delta^{(3)}(\bm{x}-\bm{a}_{i})\delta^{(3)}(\bm{y}-\bm{a}_{j})\tilde{G}_{\mathrm{R}}^{(\varsigma_h)}(\bm{a}_i,\bm{a}_j;\omega)
\end{equation}
describes at the microscopic level the spatially inhomogeneous nature of the dielectric. Its retardation nature naturally emerges and is clearly seen from the presence of the retarded Green's functions $\tilde{G}_{\mathrm{R}}^{(\varsigma_h)}(\bm{x},\bm{y};\omega)$ of the dielectric atoms at each atom's location $\bm{a}_j$. Thus, we do not need to presume this property a priori.

If the dielectric is made up of identical atoms and the nonlocal mutual influences between the dielectric atoms are ignored, then the explicit form of $\tilde{G}_{\mathrm{R}}^{(\varsigma_h)}(\bm{x},\bm{y};\omega)$  is in fact independent of the location,
\begin{equation}\label{E:krjfs}
    \tilde{G}_{\mathrm{R}}^{(\varsigma_h)}(\bm{x},\bm{y};\omega)=\frac{1}{m(-\omega^2-i\,2\gamma_{\varsigma}\omega+\varpi^2_{\varsigma})}\,.
\end{equation}
Eq.~\eqref{E:krjfs} recovers the standard result of the Lorentz model for the dispersive, absorptive dielectric medium~\cite{Ja98}, and it (or the generalization based on Eq.~\eqref{E:eudwisd}) can be easily extended to the dielectric comprising assorted species of atoms in specific arrangement. However, the task is expected to be intimidating. In particular, considering an extremely large number of atoms in the dielectric medium, the local dielectric response, as well as the local microscopic field obtained from Eq.~\eqref{E:ndeiee}, can be very complicated and tempestuous. To make contact with the macroscopic description of the dielectric-modified ambient field, certain averaging procedures as outlined in~\cite{Ja98,Ru70, Ro71, Ro73,Ch10,Sa19} must be introduced.

Suppose we have introduced an appropriate protocol for spatial averaging, with a smearing scale  which is shorter than that of our interest but much greater than the inter-atomic distances. Then the susceptibility $\chi_e$ or the dielectric function $\varepsilon$ would reduce essentially to a function of frequency only, while the remaining position dependence merely manifested in the form of an indicator function, defining the volume of the dielectric body. The same applies to the polarizable source $\tilde{\hat{\varrho}}(\bm{x},\omega)$. We then arrive at a wave equation of the macroscopic field modified by the dielectric medium
\begin{align}\label{E:rrtee}
    \bm{\nabla}^2\tilde{\hat{\phi}}(\bm{x},\omega)+\omega^2\varepsilon(\bm{x},\omega)\,\tilde{\hat{\phi}}(\bm{x},\omega)=-\tilde{\hat{\varrho}}(\bm{x},\omega)\,,
\end{align}
whose solution in the time domain in general will take the form
\begin{equation}\label{E:oghsdww}
    \hat{\phi}(\bm{x},t)=\hat{\phi}_h(\bm{x},t)+\int\!d^4x'\;G_{\mathrm{R}}^{(\phi_h)}(\bm{x},t;\bm{x}',t')\,\hat{\varrho}(\bm{x},t')\,.
\end{equation}
The first term, the homogeneous solution $\hat{\phi}_h(\bm{x},t)$, is the solution we will obtain in the absence of the sources on the right hand side of Eq.~\eqref{E:rrtee}. It is just the ``free field'' in the presence of the dielectric. (We have added quotation marks on the term ``free field" because according to conventional electrodynamics, such a field is free in the sense that it does not interact with any external charge. However, in the present framework of interacting open systems, this field is the homogeneous component of an interacting field, the consequence of the interaction between the field and the dielectric atoms.) The second term on the right hand side of Eq.~\eqref{E:oghsdww} results from the source $\hat{\varrho}(\bm{x},t')$, which can be viewed as the radiation field generated by the bounded polarizable sources in the dielectric medium.

\subsection{Internal dynamics of the system atom} 

We now reach the top tier of the nested set of graded effective actions taking into account the coarse-grained effects of the lower tiers, namely, the internal degrees of freedom of the system atoms. After we substitute the solution~\eqref{E:oghsdww} into Eq.~\eqref{E:nbks4}, we obtain the quantum Langevin equation for the idfs for the $n^{\text{th}}$ atom located at $\bm{z}_n$, referred to earlier as the internal dynamics of the  $N$-atoms system: 
\begin{align}\label{E:eerbesds}
    M\,\ddot{\hat{\chi}}^{(n)}(t)+M\Omega^2\,\hat{\chi}^{(n)}(t)+&g_{\textsc{f}}^2\sum_{n'}\int^t\!\!dt'\;\partial_tG_{\mathrm{R}}^{(\phi_h)}(\bm{z}_n,t;\bm{z}_{n'},t')\,\dot{\hat{\chi}}^{(n')}(t')\notag\\
    &=g_{\textsc{f}}\,\dot{\hat{\phi}}_h(\bm{z}_n,t)+g_{\textsc{f}}\int^t\!\!dt'\!\int\!d^3\bm{x}'\;\partial_tG_{\mathrm{R}}^{(\phi_h)}(\bm{z}_n,t;\bm{x}',t')\,\hat{\varrho}(\bm{x},t')
\end{align}
where $g_{\textsc{f}}$ measures the coupling strength between the idfs of the atoms with the dielectric modified ambient quantum field, and  $\hat{\varrho}(\bm{x},t)$ is given by the right hand side of \eqref{E:beoitue}. Note that Eq.~\eqref{E:eerbesds}  has a form similar to the typical Langevin equation describing the stochastic motion of a quantum Brownian oscillator, except for the additional term on the right hand side due to $\hat{\varrho}(\bm{x},t)$. To better understand its meaning, we first observe that the term $g_{\textsc{f}}\,\dot{\hat{\phi}}_h(\bm{z}_n,t)$ gives the quantum noise of the ambient homogeneous field in the presence of the dielectric. The second term, as a consequence of the retarded Green's function~\cite{Ja98, HH19c, BHH23}, describes the quantum noise associated with the radiation fields emitted by the intrinsic dipoles of the dielectric atoms and their induced dipoles by the thermal bath, as mentioned above. In principle, there should be a contribution related to the radiation field of the idfs of the $N$ system atoms. Indeed there is -- it lies on the left hand side of Eq.~\eqref{E:eerbesds}. As per what we have elaborated for the dielectric atoms, this term will cause the renormalization of parameters in the equation of motion for $\hat{\chi}^{(n)}$ and the mutual causal influences amongst the $N$ atoms; the latter will likewise include similar influences from the image atoms with time delays to the interface of the dielectric bulk.

We can put these equations in a more compact matrix form as follows,
\begin{equation}\label{E:eruskjhfs}
	M\ddot{\bm{X}}(t)+M\Omega^{2}\,\bm{X}(t)+g_{\textsc{f}}^{2}\int_{0}^{t}dt'\;\partial_t\bm{G}_{\mathrm{R}}^{(\phi_h)}(t,t')\cdot\dot{\bm{X}}(t')=g_{\textsc{f}}\,\dot{\bm{\xi}}^{(\phi)}(t)\,,
\end{equation}
where $\bm{X}$, $\bm{\xi}$ are the $N$-dimensional column vectors, with the $n^{\text{th}}$ element defined by $\bm{X}_{n}=\chi^{(n)}$, and
\begin{align}
    \bm{\xi}_{n}(t)=\hat{\phi}_h(\bm{z}_{n},t)+\int^t\!\!dt'\!\int\!d^3\bm{x}'\;G_{\mathrm{R}}^{(\phi_h)}(\bm{z}_n,t;\bm{x}',t')\,\hat{\varrho}(\bm{x},t')\,. 
\end{align}
The Green tensor $\bm{G}_{\mathrm{R}}^{(\phi_h)}(t,t')$ is represented by a symmetric matrix $N\times N$ with elements $\bigl[\bm{G}_{\mathrm{R}}^{(\phi_h)}\bigr]_{mn}(t,t')$ that denote $G_{\mathrm{R}}^{(\phi_h)}(\bm{z}_{m},t;\bm{z}_{n},t')$.

Formally, the general solution to Eq.~\eqref{E:eruskjhfs} can be found if we carry out the Laplace transformation of Eq.~\eqref{E:eruskjhfs} on a time interval $0\leq t<\infty$,
\begin{align}
	\Bigl[M\bigl(s^{2}+\Omega^{2}\bigr)\bm{I}_{N}+g^{2}_{\textsc{f}}\,s^2\,\overline{\bm{G}}^{(\phi_h)}_{\mathrm{R}}(s)\Bigr]\cdot\overline{\bm{X}}(s)&=g_{\textsc{f}}\,\Bigl[s\,\overline{\bm{\xi}}(s)-\bm{\xi}(0)\Bigr]+\Bigl[Ms\,\bm{I}_{N}+g^{2}_{\textsc{f}}\,s\,\overline{\bm{G}}^{(\phi_h)}_{\mathrm{R}}(s)\Bigr]\cdot\bm{X}(0)\notag\\
 &\qquad\qquad\qquad\qquad\qquad\qquad+M\dot{\bm{X}}(0)\,,
\end{align}
such that
\begin{align}\label{E:gbdiur}
   \overline{\bm{X}}(s)&=g_{\textsc{f}}\,\overline{\bm{G}}_{\mathrm{R}}^{(\chi_h)}(s)\cdot\Bigl[s\,\overline{\bm{\xi}}(s)-\bm{\xi}(0)\Bigr]+\cdots\,,
\end{align}
where $\cdots$ represents the contribution that depends on the initial conditions, and the retarded Green's function of $\chi^{(n)}$ in the frequency domain is given by
\begin{align}\label{E:oiuehbd}
    \overline{\bm{G}}_{\mathrm{R}}^{(\chi_h)}(s)&=\frac{1}{M}\,\Bigl[\bigl(s^{2}+\Omega^{2}\bigr)\bm{I}_{N}-\frac{g^{2}_{\textsc{f}}}{M}\,s^2\,\overline{\bm{G}}^{(\phi_h)}_{\mathrm{R}}(s)\Bigr]^{-1}\,.
\end{align}
The Laplace transformation of a function $f(t)$ is defined by
\begin{align}
	\int_{0}^{\infty}\!dt\;e^{-st}\,f(t)&=\overline{f}(s)\,,
\end{align}
such that
\begin{equation}
    \int_{0}^{\infty}\!dt\;e^{-st}\,\dot{f}(t)=s\,\overline{f}(s)-f(0)\,.
\end{equation}    
Note that for a retarded kernel $G_{\textrm{R}}(t)$ we have $G_{\textrm{R}}(0)=0$ by construction. Here $\bm{I}_{N}$ is the $N$-dimensional identity matrix. The inverse Laplace transformation then brings back the solution $\bm{X}(t)$.

\subsubsection{Subtleties in third time derivative terms in EOM of reduced system}

The dynamics of $\bm{X}(t)$, given by the formal solution derived from Eq.~\eqref{E:oiuehbd}, is much subtler than it appears. Let us review Eq.~\eqref{E:eerbesds}. As will be shown later, the Green's function of the ambient field outside the dielectric can be decomposed into two contributions: 1) free field in the unbounded space in the absence of the dielectric and 2) modifications due to a nearby dielectric medium. Thus, the nonlocal expression on the left hand side in fact contains four distinct contributions: 1) the local self-force on the $n^{\text{th}}$ atom, and causal nonMarkovian influences from 2) the image of the $n^{\text{th}}$ atom itself, from 3) the other $N-1$ atoms and from 4)  the images of the other $N-1$ atoms. Additional complexity arises from the local self force.

 Let us focus for the moment on issues of the self force, and write the left hand side of Eq.~\eqref{E:eerbesds} as
\begin{equation}\label{E:bieura}
     M\,\ddot{\hat{\chi}}^{(n)}(t)+M\Omega^2\,\hat{\chi}^{(n)}(t)+g_{\textsc{f}}^2\int^t\!\!dt'\;\partial_tG_{\mathrm{R,0}}^{(\phi_h)}(\bm{z}_n,t;\bm{z}_{n},t')\,\dot{\hat{\chi}}^{(n)}(t')+\cdots=g_{\textsc{f}}\,\dot{\hat{\phi}}_h(\bm{z}_n,t)\,,
\end{equation}
where $\cdots$ represents the other three nonMarkovian contributions of the nonlocal term. If the massless ambient field is coupled to a point charge, then there is  no apparent cutoff in its frequency spectrum, so the retarded Green's function $G_{\mathrm{R},0}^{(\phi_h)}(\bm{z}_n,t;\bm{z}_{n},t')$ of the $\phi_h$ in the unbounded space in the absence of the dielectric has the form
\begin{equation}\label{E:itrrrt}
    G_{\mathrm{R},0}^{(\phi_h)}(\bm{z}_n,t;\bm{z}_{n},t')=-\frac{\delta(t-t')}{2\pi}\,\partial_t\delta(t-t')\,.
\end{equation}
Then Eq.~\eqref{E:bieura} becomes
\begin{equation}\label{E:oeuhd}
     M_{\textsc{a}}^{\vphantom{2}}\,\ddot{\hat{\chi}}^{(n)}(t)+M_{\textsc{a}}^{\vphantom{2}}\Omega_{\textsc{a}}^2\,\hat{\chi}^{(n)}(t)-\dfrac{g^2_{\textsc{f}}}{4\pi}\,\dddot{\hat{\chi}}^{(n)}(t)+\cdots=g_{\textsc{f}}\,\dot{\hat{\phi}}_h(\bm{z}_n,t)\,.
\end{equation}    
where $M_{\textsc{a}}$ is the effective mass of the harmonic oscillator depicting the idf of the system atom, which receives a correction $\delta M$ from the nonlocal term in \eqref{E:bieura}, and $\Omega_{\textsc{a}}$ is the corresponding physical frequency that accompanies the re-definition of the mass. The third term is the source of complexity:  the equation of motion has a third-order time derivative, akin to that showing radiative reaction. If we literally solve such an equation of motion, we in general have three independent solutions, two of which in this case is oscillatory in time with decaying amplitude. However, owing to a tiny time scale $\tau=\dfrac{g^2_{\textsc{f}}}{4\pi M_{\textsc{a}}}$ in Eq.~\eqref{E:oeuhd}, which amounts to the classical radius of a classical charge, the third  solution grows exponentially with time, which is considered unphysical. This is the familiar `runaway' solution issue in radiation theory. The associated pathologies and proposed remedies have been well discussed in~\cite{Ja98,Ro90,Ya06,Sp04,HH22}. In the context of nonrelativistic point charge, the simplest cure is to discard the runaway solution, so that the remaining solutions describe motion similar to a driven damped oscillator. Another common practice is to implement an order reduction~\cite{Ja98,Ro90,Ya06,Sp04}, writing the third-order time derivative in \eqref{E:bieura} as a first-order time derivative,
\begin{equation}\label{E:bdjuer}
    M_{\textsc{a}}^{\vphantom{2}}\,\ddot{\hat{\chi}}^{(n)}(t)+M_{\textsc{a}}^{\vphantom{2}}\Omega_{\textsc{a}}^2\,\hat{\chi}^{(n)}(t)+M_{\textsc{a}}^{\vphantom{2}}\Omega_{\textsc{a}}^2\,\tau_{\textsc{a}}\,\dot{\hat{\chi}}^{(n)}(t)+\cdots=g_{\textsc{f}}\,\dot{\hat{\phi}}_h(\bm{z}_n,t)\,.
\end{equation}    
In so doing, Eq.~\eqref{E:bdjuer} has a form similar to Eq.~\eqref{E:bqwrr} with an effective damping constant $\gamma_{\textsc{a}}=\Omega_{\textsc{a}}^2\tau_{\textsc{a}}/2$ and $\tau_{\textsc{a}}$ is the effective radius of the classical charge. Note that here $\tau_{\textsc{a}}$ in the order-reduced Eq.~\eqref{E:bdjuer} can  in principle be different from $\tau$ in Eq.~\eqref{E:oeuhd}, depending on the choice of frequency used in the order-reduction protocol. This new equation is often called the nonrelativistic Landau-Lifshitz equation for a point charge. It does not contain any runaway mode, so its solution is well behaved and damps with time as well.

\subsubsection{Consistent order reduction for EOM in Gaussian open systems}

In the above order-reduction protocol, only the dissipation backaction {from the ``environment"} on the idf of the system atoms is treated, but the accompanying fluctuation backaction is left untouched. This protocol, though popular,  may undermine the self-consistency requirement because in the framework of Gaussian open systems, both backactions are on equal footing. They are not mutually independent, but connected by the statistical properties of linear {environmental} quantum field, that is, the fluctuation-dissipation relations. Hence the preferential treatment on the dissipative backreaction alone may inadvertently modify the relaxation behavior of the internal degrees of freedom $\chi^{(n)}$ of the system atom, as well as the delicate dynamical balances with the other three nonMarkovian influences in Eq.~\eqref{E:eerbesds} via the correlation-propagation relations~\cite{HH19a}.

To remedy this insufficiency, we observe that apart from the fluctuation backactions, the order-reduced equation of motion has a form similar to that for the Brownian motion, where the dynamical variable of the system is coupled to the field, not the derivative of the field. It suggests that if the order reduction is properly applied to the fluctuation backactions on the dipole, the latter will also reduce to its counterpart for the Brownian motion. Following this line of reasoning, we arrive at Eq.~\eqref{E:bekoer} in Appendix~\ref{App:bebeirie} or Eq.~\eqref{E:feireer} below, where both backactions on the dipole are order reduced. The dipole system dynamics described by such an equation of motion then has a well-behaved asymptotic equilibrium state. Moreover, the consistency manifests in a simple yet familiar form of the fluctuation-dissipation relation for the dipole in Eq.~\eqref{E:soieuro}. 
%Of course, here we only propose a plausible prescription to consistently implement the order reduction for a Gaussian open system. 

Using the order reduction prescription proposed in Appendix~\ref{App:bebeirie}, we can cast Eq.~\eqref{E:eruskjhfs} into
\begin{equation}\label{E:feireer}
	M_{\textsc{a}}\ddot{\bm{X}}(t)+M_{\textsc{a}}\Omega^{2}\,\bm{X}(t)-g'^{2}_{\textsc{f}}\int_{0}^{t}dt'\;\bm{G}_{\mathrm{R}}^{(\phi_h)}(t,t')\cdot\bm{X}(t')=g'_{\textsc{f}}\,\bm{\xi}^{(\phi)}(t)\,,
\end{equation}
where $g'_{\textsc{f}}=g_{\textsc{f}}\Omega$ is the effective/physical coupling strength. {Eq.~\eqref{E:feireer} will be the starting point to compute the ingredients for constructing the entanglement measures, as outlined below.}

Now that we have all the necessary ingredients ready,  we end this long section with a general remark:  
For linear systems, the quantum Langevin equation approach usually offers a more transparent picture of the physics involved.  We shall use it to explore the complexities laden in the graded interacting systems and their many physical manifestations. In what follows, we will assume that the dielectric has a finite size but its linear extension is much greater than the length scale of our interest, so that in the vicinity of the dielectric, the internal dynamics of the neutral atoms has finite relaxation time scales~\cite{HH18a}, but we can still neglect the finite-size, edge effects of the dielectric. We further assume the values of $g_{\textsc{a}}$, $g_{\textsc{m}}$, and $g_{\textsc{b}}$, in particular the latter two, are very small such that the contributions from the dielectric dipoles induced by the bath field are negligible within suitable time scales. Under these conditions, we can focus on the effects of the  dielectric-modified  macroscopic ambient field on the neutral atom outside the dielectric.

\section{Covariance Matrix and Entanglement Measures}\label{S:eeuidf}

In general, the elements of the density matrix of a Gaussian system can be fully characterized by the covariance matrix of the system. If the system contains $N$ pairs of canonical variable $(\chi^{(n)},p^{(n)})$, where $p^{(n)}$ is the momentum conjugated to $\chi^{(n)}$ with $n=1$, $\cdots$, $N$, then the covariance matrix is a $2N\times 2N$ matrix, defined by
\begin{equation}
	\bm{V}_{ij}=\frac{1}{2}\,\langle\{O_{i},\,O_{j}\}\rangle\,,
\end{equation}
in which $\bm{O}=(\chi^{(1)},p^{(1)},\ldots,\chi^{(N)},p^{(N)})$. Here without loss of generality, we have assumed $\langle\bm{O}\rangle=0$. This is always possible by re-defining $\bm{O}$ by a shift $\bm{O}-\langle\bm{O}\rangle$. The quantum expectation value $\langle\cdots\rangle$ is defined with respect to the initial state of the system.

The internal degrees of the freedom $\bm{X}$ of the system atoms, described by the order-reduced Langevin equation Eq.~\eqref{E:feireer}, has well behaved relaxation dynamics. It will tend to an asymptotic equilibrium state under pretty broad conditions~\cite{Ve11,HH18a}. There the elements of the covariance matrix evolve to time-independent values, given by
\begin{align}
	\bm{V}_{XX}(\infty)&=\operatorname{Im}\int_{-\infty}^{\infty}\frac{d\omega}{2\pi}\;\coth\frac{\beta\omega}{2}\,\tilde{\bm{G}}_{\mathrm{R}}^{(\chi_h)}(\omega)\,,\label{E:wutbs1}\\
	\bm{V}_{PP}(\infty)&=M_{\textsc{a}}^2\operatorname{Im}\int_{-\infty}^{\infty}\frac{d\omega}{2\pi}\;\omega^{2}\coth\frac{\beta\omega}{2}\,\tilde{\bm{G}}_{\mathrm{R}}^{(\chi_h)}(\omega)\,,\label{E:wutbs2}\\
	\bm{V}_{XP}(\infty)&=0\,,\label{E:wutbs3}
\end{align}
in terms of $\bm{G}_{\mathrm{R}}^{(\chi_h)}$ in momentum space,
\begin{equation}
    \tilde{\bm{G}}_{\mathrm{R}}^{(\chi_h)}(\omega)=\biggl[M_{\textsc{a}}\Bigl(-\omega^2+\Omega^2\Bigr)\,\bm{I}_N-g'^2_{\textsc{f}}\,\tilde{\bm{G}}_{\mathrm{R}}^{(\phi_h)}(\omega)\biggr]^{-1}\,,
\end{equation}
where a reduction of order has been implemented. {The derivations of Eqs.~\eqref{E:wutbs1}--\eqref{E:wutbs3} are sketched in Appendix~\ref{App:ebfsjdbfs}.} Here we conjecture that in the presence of the absorptive dielectric, such a final equilibrium state exists if, under the assumption we made in the beginning of this section, the state of the macroscopic ambient field is already a thermal state of temperature $\beta^{-1}$ before the dipoles of the system atoms are coupled to the field. As can be seen, all the information of the modified field is wrapped in $\tilde{\bm{G}}^{(\phi_h)}_{\mathrm{R}}(\omega)$ inside the expression of $\tilde{\bm{G}}^{(\chi_h)}_{\mathrm{R}}(\omega)$, the former in turn depends on the permittivity $\varepsilon$.

As an example, suppose that we have two polarizable system atoms located at the fixed positions $\bm{z}_{n}$, $(n=1$, $2)$ and with the same mass $M_{\textsc{a}}$ with idf variables $\chi^{(n)}$. If the ambient field is initially in its vacuum state, i.e, $\beta\to\infty$, then we can write down explicitly their formal {late-time} expressions as follows 
\begin{align*}
	V_{XX}(\infty)=\frac{1}{M_{\textsc{a}}}\operatorname{Im}\!\int_{0}^{\infty}\!\frac{d\omega}{\pi}\begin{pmatrix}
-\omega^{2}+\Omega^{2}-\dfrac{g'^{2}_{\textsc{f}}}{M_{\textsc{a}}}\,\tilde{G}_{\mathrm{R}}^{(\phi_h)}(\bm{z}_{1},\bm{z}_{1};\omega)
& -\omega^{2}+\Omega^{2}-\dfrac{g'^{2}_{\textsc{f}}}{M_{\textsc{a}}}\,\tilde{G}_{\mathrm{R}}^{(\phi_h)}(\bm{z}_{1},\bm{z}_{2};\omega)\vspace{8pt}\\
-\omega^{2}+\Omega^{2}-\dfrac{g'^{2}_{\textsc{f}}}{M_{\textsc{a}}}\,\tilde{G}_{\mathrm{R}}^{(\phi_h)}(\bm{z}_{2},\bm{z}_{1};\omega) & -\omega^{2}+\Omega^{2}-\dfrac{g'^{2}_{\textsc{f}}}{M_{\textsc{a}}}\,\tilde{G}_{\mathrm{R}}^{(\phi_h)}(\bm{z}_{2},\bm{z}_{2};\omega)\end{pmatrix}^{-1}\,.
\end{align*}
This can easily be generalized to an arbitrary number of atoms.

In addition to the fact that a Gaussian state can be characterized by the covariance matrix elements, they are also the building blocks of many physical observables in the Gaussian systems, from the Robertson-Schrodinger uncertainty relation~\cite{DR22}, to the thermodynamic functions~\cite{HH18a,HH21} in nonequilibrium quantum thermodynamics,  such as the free energy, entropy and so on,  and for quantum information, entanglement measures like linear entropy, purity for  pure-state entanglement~\cite{LH09,HH15b,Pe96,AS04,AS05,PI03} or negativity, Peres-Horodecki-Simon separability criterion for  mixed-state entanglement~\cite{LH09,HH15b,HH96,AI05,Pl05,Si00,AR14,Ad07,AH23}. They can  even be used to express the nonequilibrium transition probability due to strong atom-field interaction when a harmonic atom is placed in a linear quantum field~\cite{HH19b}.

Here we will use purity $\mu$ as an illustrative example. When a neutral atom initially prepared in a pure state is coupled to the surrounding quantum field in its initial vacuum state, their interaction or mutual influence can put them into an entangled Gaussian state. When this happens, the purity of the atomic state will drop below unity because after we trace over the field sector, the reduced density matrix of the atom becomes mixed. Thus purity can serve as a measure for   pure-state entangled systems. The expression of purity takes a particularly simple form after the internal dynamics of the atom asymptotically approaches an equilibrium state as $t\to\infty$, where the covariance matrix element $V_{XP}$ vanishes. The purity $\mu$ in this asymptotic equilibrium state is given by
\begin{align}\label{E:psjwrb}
	\mu=\operatorname{Tr}\{\rho_{s}^{2}\}=\frac{1}{2\sqrt{\det\bm{V}}}=\frac{1}{2\sqrt{V_{XX}(\infty)V_{PP}(\infty)}}\,.
\end{align}
Note that in the current configuration, $\nu^2=\det\bm{V}$ also gives the Robertson-Schr\"odinger function. Since it is bounded below by $1/4$, this is a useful criterion for checking the consistency of numerical computations.

The von Neumann entropy is another popular measure used in pure-state entanglement. Following the same line of argument, when the system atom is entangled with the ambient field via the dipole-field interaction, the atom is in a mixed state, and its von Neumann entropy becomes nonzero. If the idf of the system atom is in a Gaussian state, the von Neumann entropy can  likewise be expressed in terms of the elements of the covariance matrix,
\begin{align}
    S_{vN}=\Bigl(\nu+\frac{1}{2}\Bigr)\,\ln\Bigl(\nu+\frac{1}{2}\Bigr)-\Bigl(\nu-\frac{1}{2}\Bigr)\,\ln\Bigl(\nu-\frac{1}{2}\Bigr)\,.
\end{align}
Then the results we will obtain later in Eqs~\eqref{E:triu1} and \eqref{E:triu2} can be used to find the leading contribution when the coupling $g'_{\textsc{f}}$ is weak. Otherwise, to deal with nonperturbative expressions such as Eqs.~\eqref{E:oeurwiw1} and \eqref{E:oeurwiw2} for the cases of finite coupling strength, we may need to resort to numerical methods.

In what follows, we will formally compute the ingredients needed in Eq.~\eqref{E:psjwrb} for the discussions of quantum entanglement between one neutral atom and the ambient quantum field modified by the dispersive, absorptive dielectric half-space nearby.

\section{Green's Functions of Quantum Field outside a Dielectric Half-space}

We consider the simplest scenario where the linear dielectric medium occupies half of the space and the initial state of the ambient quantum field is in a vacuum state before the neutral atom is introduced and coupled to.

%\subsection{Quantum Correlations of Atom Influenced by Half-Space Medium }
The retarded Green's function of the dielectric-modified macro field in the frequency space satisfies the equation
\begin{align}\label{E:ernbdkbhs}
	\Bigl[\bm{\nabla}^{2}+\omega^{2}\,\varepsilon(\bm{x},\omega)\Bigr]\,\widetilde{G}_{\mathrm{R}}^{(\phi_h)}(\bm{x},\bm{x}';\omega)=-\delta^{(3)}(\bm{x},\bm{x}')\,,
\end{align}
with the dielectric function $\varepsilon(\bm{x},\omega)$, after suitable spatial averaging, given by
\begin{align}\label{E:bbyuee}
	\varepsilon(\bm{x},\omega)&=1+g_{\textsc{m}}^{2}\,n\,\bm{1}_{V}(\bm{x})\,\widetilde{G}^{(\varsigma_h)}_{\mathrm{R}}(\omega)\,,&&\text{and}&\widetilde{G}^{(\varsigma_h)}_{\mathrm{R}}(\omega)&=\frac{1}{m[w_{\varsigma}^{2}-(\omega+i\,\gamma_{\varsigma})^{2}]}\,,
\end{align}
where $w_{\varsigma}$ is $w_{\varsigma}=\sqrt{\varpi_{\varsigma}^2-\gamma_{\varsigma}^{2}}$, defined in \eqref{E:hgurte}, $\bm{1}_{V}(\bm{x})$ is the indicator function
\begin{equation}
    \bm{1}_{V}(\bm{x})=\begin{cases}
        1\,,&\bm{x}\in V\,,\\
        0\,,&\bm{x}\notin V\,.
    \end{cases}
\end{equation}
and $V$ represents the volume of the dielectric half space. We suppose that the dielectric atoms are arranged in square lattices, of unit volume $a^3$, and each unit lattice contains only one scalar charge $g_{\textsc{f}}$,  so the charge density $n$ is  given by $1/a^3$. 

As is well-known~\cite{Ja98}, in the high frequency limit, this dielectric behaves like 
\begin{align}
	\lim_{\omega\to\infty}\varepsilon(\bm{x},\omega)&\simeq1-\frac{g_{\textsc{m}}^{2}}{m\omega^{2}a^{3}}\,,
\end{align}
where a ``plasma frequency'' $\omega_{p}$ for this dielectric model can be identified by
\begin{equation}\label{E:wwruhe}
	\omega_{p}^{2}=\frac{g_{\textsc{m}}^{2}}{ma^{3}}\,.
\end{equation}
For the modes of the ambient field whose frequencies are higher than this plasma frequency~\eqref{E:wwruhe}, the constituent atoms in the dielectric hardly respond to the quantum fluctuations of these modes, so the dielectric medium appears essentially transparent to them. The wavelength $\lambda_{p}$ corresponding to the plasma frequency is of the order
\begin{equation}
	\lambda_{p}\sim \sqrt{\frac{\pi a^3}{2\gamma_{\textsc{m}}}}\,,
\end{equation}
where the damping constant $\gamma_{\textsc{m}}$, corresponding to the coupling constant $g_{\textsc{m}}$, is defined by $\gamma_{\textsc{m}}=g_{\textsc{m}}^{2}/8\pi m$. For weak coupling, the lattice spacing and the plasma wavelength are quite distinct scales and for typical metals,  $\lambda_{p}\gg a$, that is, $\omega_{p}\ll a^{-1}$. It implies that for field modes up to this plasma frequency, the dielectric medium can be treated as a continuum because the typical wavelengths of the field are much larger than the lattice spacing, which is of the same order of magnitude as the atom size, unless the dielectric medium is extremely dilute such that the inter-atomic spacing is much greater than the atom size. Hence we can indeed ignore the discrete nature of the lattice and express the equations for the Green's functions of the field in terms of a continuous (or at least a piece-wise continuous) permittivity function in space. Under a suitable spatial averaging process, a microscopic dielectric function can thus be cast into a macroscopic one in Eq.~\eqref{E:bbyuee}.

In this paper, we adopt the simplest possible configuration of the medium. We assume that the dielectric medium is {macroscopically} spatially-homogeneous {after the aforementioned spatial average has been carried out,} and occupies half of the space, so the dielectric function $\varepsilon(\bm{x},\omega)$ becomes
\begin{equation}
	\varepsilon(\bm{x},\omega)=\begin{cases}
										1\,,&\text{in vacuum, $z>0$}\,,\\
										\varepsilon(\omega)\,,&\text{in the medium, $z<0$}\,.
								   \end{cases}
\end{equation}
Therefore in vacuum the equation \eqref{E:ernbdkbhs} reduces to
\begin{align}
	\Bigl[\bm{\nabla}^{2}+\omega^{2}\Bigr]\,\widetilde{G}_{\mathrm{R}}^{(\phi_h)}(\bm{x},\bm{x}';\omega)&=-\delta^{(3)}(\bm{x},\bm{x}')\,,&&\text{(vacuum, $z>0$)}\,,
\intertext{and within the medium we have}
	\Bigl[\bm{\nabla}^{2}+\omega^{2}\varepsilon(\omega)\Bigr]\,\widetilde{G}_{\mathrm{R}}^{(\phi_h)}(\bm{x},\bm{x}';\omega)&=-\delta^{(3)}(\bm{x},\bm{x}')\,,&&\text{(medium, $z<0$)}\,,
\end{align}
with 
\begin{equation}
	\varepsilon(\omega)=1+\frac{\omega_{p}^{2}}{w_{\varsigma}^{2}-(\omega+i\,\gamma_{\varsigma})^{2}}\,.
\end{equation}
The boundary conditions we impose are such that the scalar wave behaves like the TE (s) mode of the electric field across the vacuum-dielectric interface. Moreover, in the coincident limit, the Green's function of the scalar field has the usual discontinuous behavior due to the presence of the delta function on the right hand side of the wave equations.

In general, if we consider the case that $\bm{x}$, $\bm{x}'$ lie outside the dielectric, that is, $z$ and $z'>0$, then the retarded Green's function of the field usually contains two physically distinct contributions; one is the retarded Green's function for the unbounded free space and the other is the correction due to the presence of the dielectric. The retarded Green's function in the unbounded free space  can be conveniently expanded as the superpositions of plane waves propagating along the  direction normal to the interface \cite{Ch95},
\begin{equation}\label{E:ionejdk}
	\widetilde{G}_{\mathrm{R},\,0}^{(\phi_h)}(\bm{x},\bm{x}';\omega)=\frac{e^{i\,\omega\lvert\bm{x}-\bm{x}'\rvert}}{4\pi\lvert\bm{x}-\bm{x}'\rvert}=\frac{i}{2}\int_{0}^{\infty}\!\frac{dk_{\rho}}{2\pi}\;\frac{k_{\rho}}{k_{z}+i\,0^{+}}\;J_{0}(k_{\rho}\rho)\,e^{i\,k_{z}\,(z-z')}
\end{equation}
in which $\bm{x}-\bm{x}'=(\rho\cos\varphi,\rho\sin\varphi,z-z')$ in the cylindrical coordinate system with $\rho^{2}=(x-x')^{2}+(y-y')^{2}$ and $\varphi$ being the azimuth angle between the projections of $\bm{x}$, $\bm{x}'$ on the $x$-$y$ plane. The additional contribution in the retarded Green's function that accounts for the dielectric medium is given by
\begin{equation}\label{E:qhreuw}
	\widetilde{G}_{\mathrm{R},\,\mathrm{M}}^{(\phi_h)}(\bm{x},\bm{x}';\omega)=\frac{i}{2}\int_{0}^{\infty}\frac{dk_{\rho}}{2\pi}\;\left(\frac{k_{z}-K_{z}}{k_{z}+K_{z}}\right)\frac{k_{\rho}}{k_{z}+i\,0^{+}}\,J_{0}(k_{\rho}\rho)\,e^{i\,k_{z}\,(z+z')}\,,
\end{equation}
with $k_{z}=\sqrt{\omega^{2}-k_{\rho}^{2}}$ and $K_{z}=\sqrt{\omega^{2}\varepsilon(\omega)-k^{2}_{\rho}}$. It essentially describes the wave reflected off the interface at $z=0$ into the vacuum regime $z>0$. Note that this component also includes the contribution from the evanescent modes. The evanescent modes occur when $\omega^{2}<k_{\rho}^{2}<\omega^{2}\varepsilon(\omega)$, if $\omega^{2}\varepsilon(\omega)>\omega^{2}$ is assumed. In this case $k_{z}$ can be purely imaginary and the amplitude of the modes will decay with the distance away from the dielectric-vacuum interface as they further travel into the vacuum. Thus this evanescent mode propagates only in the directions parallel to the interface.

The contribution \eqref{E:qhreuw} may be interpreted as the retarded field produced by a dressed image point source  located at the image point $\overline{\bm{x}}'$ of $\bm{x}'$. However, this interpretation may not be  useful when the ``index of refraction'' or the dielectric function is mode-dependent.

\section{One Atom in a Quantum Field above a Dielectric Half-space}

In the remainder of this paper we shall specialize to one system atom, so that we can understand the  essential physics of atom-field-medium interaction with minimal computational burden. If we place only one atom at a distance $\ell$ above the half-space dielectric medium, then $\bm{z}_1=\bm{x}=\bm{x}'$, so the retarded Green's function $\widetilde{G}_{\mathrm{R}}^{(\chi_h)}(\omega)$ of the atom in Eq.~\eqref{E:feireer} reduces to
\begin{equation}\label{E:omernjke}
	\widetilde{G}_{\mathrm{R}}^{(\chi_h)}(\omega)=\frac{1}{M_{\textsc{a}}}\,\Bigl[-\omega^{2}+\Omega^{2}_{\textsc{a}}-i\,2\gamma'_{\textsc{a}}\omega-\frac{g'^{2}_{\textsc{f}}}{M_{\textsc{a}}}\,\widetilde{G}_{\mathrm{R},\,\mathrm{M}}^{(\phi_h)}(\omega)\Bigr]^{-1}\,,
\end{equation}
in the frequency space, where $\widetilde{G}_{\mathrm{R},\,M}^{(\phi_h)}(\omega)$ is given by \eqref{E:qhreuw} with $\rho=0$ and $z=z'=\ell$, i.e., $\bm{x}=\bm{x}'$. Contrast this with the unbounded free Green's function of the scalar field $G_{\mathrm{R},\,0}^{(\phi_h)}$. When evaluated in the coincident limit in the coordinate space, the latter gives the damping constant, defined by $\gamma'_{\textsc{a}}=g'^{2}_{\textsc{f}}/8\pi M_{\textsc{a}}$, and renormalizes the bare natural frequency to its physical value $\Omega_{\textsc{a}}$.

\subsection{Late-time Covariances of the Atom}

Given $\widetilde{G}_{\mathrm{R}}^{(\chi_h)}(\omega)$ we can evaluate the late-time covariances of the canonical variables of the atom, by eqs.~\eqref{E:wutbs1}--\eqref{E:wutbs3}. Thus we find that the uncertainty in $\chi(t)$ at late time is formally given by
\begin{align}
	V_{XX}(\infty)&=\operatorname{Im}\int_{0}^{\infty}\!\frac{d\omega}{\pi}\;\widetilde{G}_{\mathrm{R},\,0}^{(\chi_h)}(\omega)\Bigl[1-g'^{2}_{\textsc{f}}\,\widetilde{G}_{\mathrm{R},\,\mathrm{M}}^{(\phi_h)}(\omega)\widetilde{G}_{\mathrm{R},\,0}^{(\chi_h)}(\omega)\Bigr]^{-1}\notag\\
	&=\operatorname{Im}\int_{0}^{\infty}\!\frac{d\omega}{\pi}\;\widetilde{G}_{\mathrm{R},\,0}^{(\chi_h)}(\omega)\sum_{n=0}^{\infty}g'^{2n}_{\textsc{f}}\,\Bigl[\widetilde{G}_{\mathrm{R},\,\mathrm{M}}^{(\phi_h)}(\omega)\widetilde{G}_{\mathrm{R},\,0}^{(\chi_h)}(\omega)\Bigr]^{n}\,,\label{E:oeurwiw1}
\end{align}
where $\widetilde{G}_{\mathrm{R},\,0}^{(\chi_h)}$ is the retarded kernel of the atom's idf variable $\chi$ in the absence of the dielectric half-space,
\begin{equation}
	\widetilde{G}_{\mathrm{R},\,0}^{(\chi_h)}=\frac{1}{M_{\textsc{a}}}\,\Bigl[-\omega^{2}+\Omega^{2}_{\textsc{a}}-i\,2\gamma'_{\textsc{a}}\omega\Bigr]^{-1}\,.
\end{equation}
The uncertainty of the corresponding canonical momentum is then given by
\begin{align}
	V_{PP}(\infty)&=M^2_{\textsc{a}}\operatorname{Im}\int_{0}^{\infty}\!\frac{d\omega}{\pi}\;\omega^{2}\,\widetilde{G}_{\mathrm{R},\,0}^{(\chi_h)}(\omega)\sum_{n=0}^{\infty}g'^{2n}_{\textsc{f}}\,\Bigl[\widetilde{G}_{\mathrm{R},\,\mathrm{M}}^{(\phi_h)}(\omega)\widetilde{G}_{\mathrm{R},\,0}^{(\chi_h)}(\omega)\Bigr]^{n}\,.\label{E:oeurwiw2}
\end{align}
The integrands in eqs.~\eqref{E:oeurwiw1} and \eqref{E:oeurwiw2} are formally the Schwinger-Dyson equation for the retarded kernel $\widetilde{G}_{\mathrm{R}}^{(\chi_h)}(\omega)$, in particular,
\begin{align}
	\widetilde{G}_{\mathrm{R}}^{(\chi_h)}(\omega)&=\widetilde{G}_{\mathrm{R},\,0}^{(\chi_h)}(\omega)\sum_{n=0}^{\infty}g'^{2n}_{\textsc{f}}\,\Bigl[\widetilde{G}_{\mathrm{R},\,\mathrm{M}}^{(\phi_h)}(\omega)\widetilde{G}_{\mathrm{R},\,0}^{(\chi_h)}(\omega)\Bigr]^{n}\notag\\
	&=\widetilde{G}_{\mathrm{R},\,0}^{(\chi_h)}(\omega)+g'^{2n}_{\textsc{f}}\,\widetilde{G}_{\mathrm{R}}^{(\chi_h)}(\omega)\widetilde{G}_{\mathrm{R},\,\mathrm{M}}^{(\phi_h)}(\omega)\widetilde{G}_{\mathrm{R},\,0}^{(\chi_h)}(\omega)\,,\label{E:ejkekjs}
\end{align}
where multiple mutual influences between the atom and the dielectric-modified scalar field are manifest.

Even for a simple dielectric, the integrals in  Eqs.~\eqref{E:oeurwiw1} and \eqref{E:oeurwiw2} cannot be obtained analytically. However, if the atom-field coupling is vanishingly weak, a condition  commonly assumed in most linear response theories, we may only keep the lowest-order contributions in $\mathcal{O}(g'^{2}_{\textsc{f}})$, and approximate Eqs.~\eqref{E:oeurwiw1} and \eqref{E:oeurwiw2} by
\begin{align}
	V_{XX}(\infty)&\simeq\operatorname{Im}\int_{0}^{\infty}\!\frac{d\omega}{\pi}\,\Bigl[\widetilde{G}_{\mathrm{R},\,0}^{(\chi_h)}(\omega)+g'^{2}_{\textsc{f}}\,\widetilde{G}_{\mathrm{R},\,0}^{(\chi_h)}(\omega)\widetilde{G}_{\mathrm{R},\,\mathrm{M}}^{(\phi_h)}(\omega)\widetilde{G}_{\mathrm{R},\,0}^{(\chi_h)}(\omega)+\mathcal{O}(g'^{4}_{\textsc{f}})\Bigr]\,,\label{E:triu1}\\
	V_{PP}(\infty)&\simeq M^2_{\textsc{a}}\operatorname{Im}\int_{0}^{\infty}\!\frac{d\omega}{\pi}\;\omega^{2}\Bigl[\widetilde{G}_{\mathrm{R},\,0}^{(\chi_h)}(\omega)+g'^{2}_{\textsc{f}}\,\widetilde{G}_{\mathrm{R},\,0}^{(\chi_h)}(\omega)\widetilde{G}_{\mathrm{R},\,\mathrm{M}}^{(\phi_h)}(\omega)\widetilde{G}_{\mathrm{R},\,0}^{(\chi_h)}(\omega)+\mathcal{O}(g'^{4}_{\textsc{f}})\Bigr]\,.\label{E:triu2}
\end{align}
Under this assumption one can use these two expressions of the covariance matrix elements of the neutral atom outside the dielectric for a perturbative  evaluation of the purity~\eqref{E:psjwrb}. For  finite values of $g'_{\textsc{f}}$ one can evaluate Eqs.~\eqref{E:oeurwiw1} and \eqref{E:oeurwiw2} by  numerical methods. The expressions developed here can be generalized to other aspects of atoms interacting with a dielectric-altered field,  as in Casimir-Poldar (CP) and dynamical CP effects,  by allowing the dielectric permittivity to be time-dependent or allowing more complex dielectric structures like the layered dielectrics~\cite{OZ24} in the context of  near-field effects, thermal transport, quantum friction or vacuum viscosity.

\subsection{Entanglement between an atom and a dielectric half-space}

\subsubsection{Perfect conductor surface}

As an appetizer we first consider a simpler configuration where the dielectric is a perfect conductor. By taking the limit $\lvert\varepsilon\rvert\to\infty$ in Eq.~\eqref{E:qhreuw}, the correction to the retarded Green's function $\widetilde{G}_{\mathrm{R},\,\mathrm{M}}^{(\phi_h)}(\bm{x},\bm{x}';\omega)$ of the ambient field due to the presence of the dielectric in Eq.~\eqref{E:qhreuw} takes a very simple form~\cite{Rong, HL06},
\begin{equation}
    \widetilde{G}_{\mathrm{R},\,\mathrm{M}}^{(\phi_h)}(\bm{x},\bm{x}';\omega)=-\frac{i}{2}\int_{0}^{\infty}\frac{dk_{\rho}}{2\pi}\;\frac{k_{\rho}}{k_{z}+i\,0^{+}}\,e^{i\,2k_{z}\ell}=-\frac{1}{8\pi\ell}\,e^{i2\omega\ell}\,.
\end{equation}
In this case, first we expect that if the atom is very far away from the conductor-vacuum interface, then the boundary effect is negligible as if it were not there. Thus, we expect that the internal dynamics of the system atom is similar to that of the Brownian oscillator in unbounded space. On the other hand, since a condition is imposed on the field  similar to that of the transverse electric field on the surface of the perfect conductor, we expect $\widetilde{G}_{\mathrm{R}}^{(\phi_h)}(\bm{x},\bm{x}';\omega)=\widetilde{G}_{\mathrm{R},\,0}^{(\phi_h)}(\bm{x},\bm{x}';\omega)+\widetilde{G}_{\mathrm{R},\,\mathrm{M}}^{(\phi_h)}(\bm{x},\bm{x}';\omega)\sim0$ when either one of $\bm{x}$, $\bm{x}'$ is very close to the boundary. Thus the idf of the system atom may behave like a undamped oscillator. This can be understood as a consequence of the damping being effectively suppressed due to very intense and rapidly back-and-forth nonMarkovian influences mediated by the ambient field.

\begin{figure}
    \centering
    \hfill
    \begin{subfigure}[b]{0.45\textwidth}
         \centering
         \includegraphics[width=0.95\linewidth]{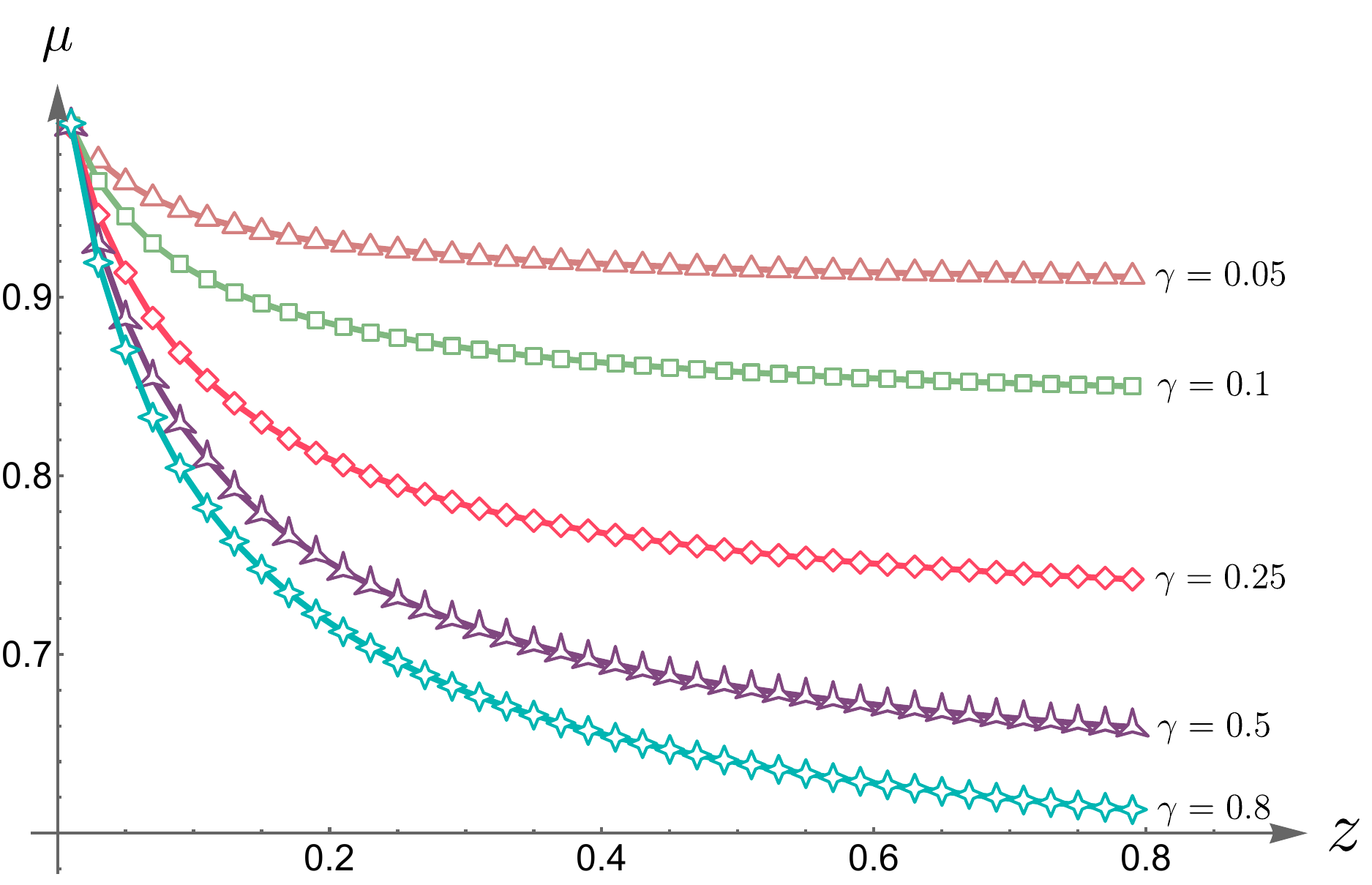}
         \caption{Variation of purity $\mu$ with the distance $z$ to the interface for selected values of the damping constant $\gamma$.}
         \label{Fi:muzPC}
    \end{subfigure}
    \hfill
    \begin{subfigure}[b]{0.45\textwidth}
         \centering
         \includegraphics[width=0.95\linewidth]{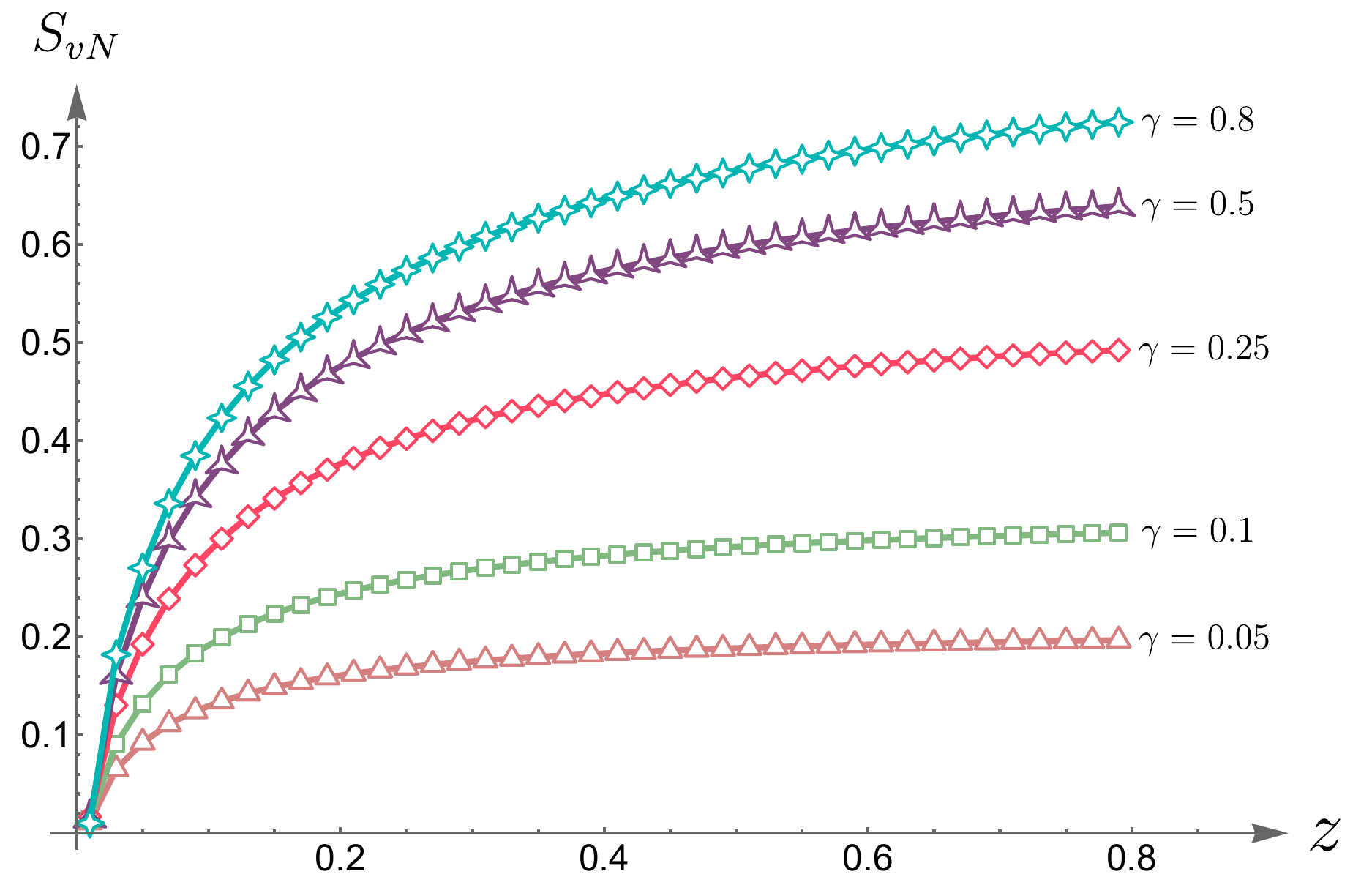}
         \caption{Variation of the von Neumann entropy $S_{vN}$ with the distance $z$ to the interface for selected values of the damping constant $\gamma$.}
         \label{Fi:svnzPC}
    \end{subfigure}
    \hfill
    \caption{The medium is a perfect conductor. Here the distance $z$ is normalized relative to the wavelength associated with the electronic transition frequency $\omega_{\textsc{a}}$, and   $\gamma$ is $\gamma'_{\textsc{a}}$ normalized with respective to $\omega_{\textsc{a}}$.}
    \label{Fi:sq}
\end{figure}
Fig.~\ref{Fi:sq} shows the results of a) purity and b) the von Neumann entropy for the extents of entanglement between the neutral atom and the dielectric-modified ambient field at late times. In the vicinity of the interface, the internal states of the system atom has high purity due to the aforementioned suppression of the effective damping (which in turn can be translated into suppression of the effective coupling, by the generic relation $\gamma'_{\textsc{a}}=g'^2_{\textsc{a}}/8\pi M_{\textsc{a}}$ between the damping constant and the coupling strength). This interpretation is clearly shown if we plot a virtual horizontal line in either subplot, then the curve with a larger choice $\gamma$ has the same degree of entanglement at smaller distance to the interface, meaning the coupling constant having been sufficiently subdued.

With increasing $z$, the final equilibrium state of the idf of the system atom becomes increasing mixed due to the atom-field interaction. Thus purity decreases, while the von Neumann entropy grows, signifying an increase in atom-field entanglement. 
%That is, it is harder to decouple the information in the atomic degree of freedom and that in the field degree of freedom. 

Fig.~\ref{Fi:sq} also shows that stronger atom-field coupling  increases the mixture of their states, such that they become  more difficult to disentangle. The results are consistent with our physical intuitions. Note that we did not push $z$ all the way to zero. In that regime, the configuration becomes considerably complicated because we cannot apply the macro field description. Even though we disregard the issues associated with the micro field description, the renormalization of the parameters for the atom's internal degrees of freedom are less well-defined because the correction due to the presence of the medium becomes significantly appreciable. In addition, it is known that close to the boundary, we have towering  (renormalized) field fluctuations, impinging on the atoms. It is not yet clear whether the reduced dynamics of the atom's idf can be meaningfully formulated under such an extreme condition.

\begin{figure}
    \centering
    \hfill
    \begin{subfigure}[b]{0.45\textwidth}
         \centering
         \includegraphics[width=0.95\linewidth]{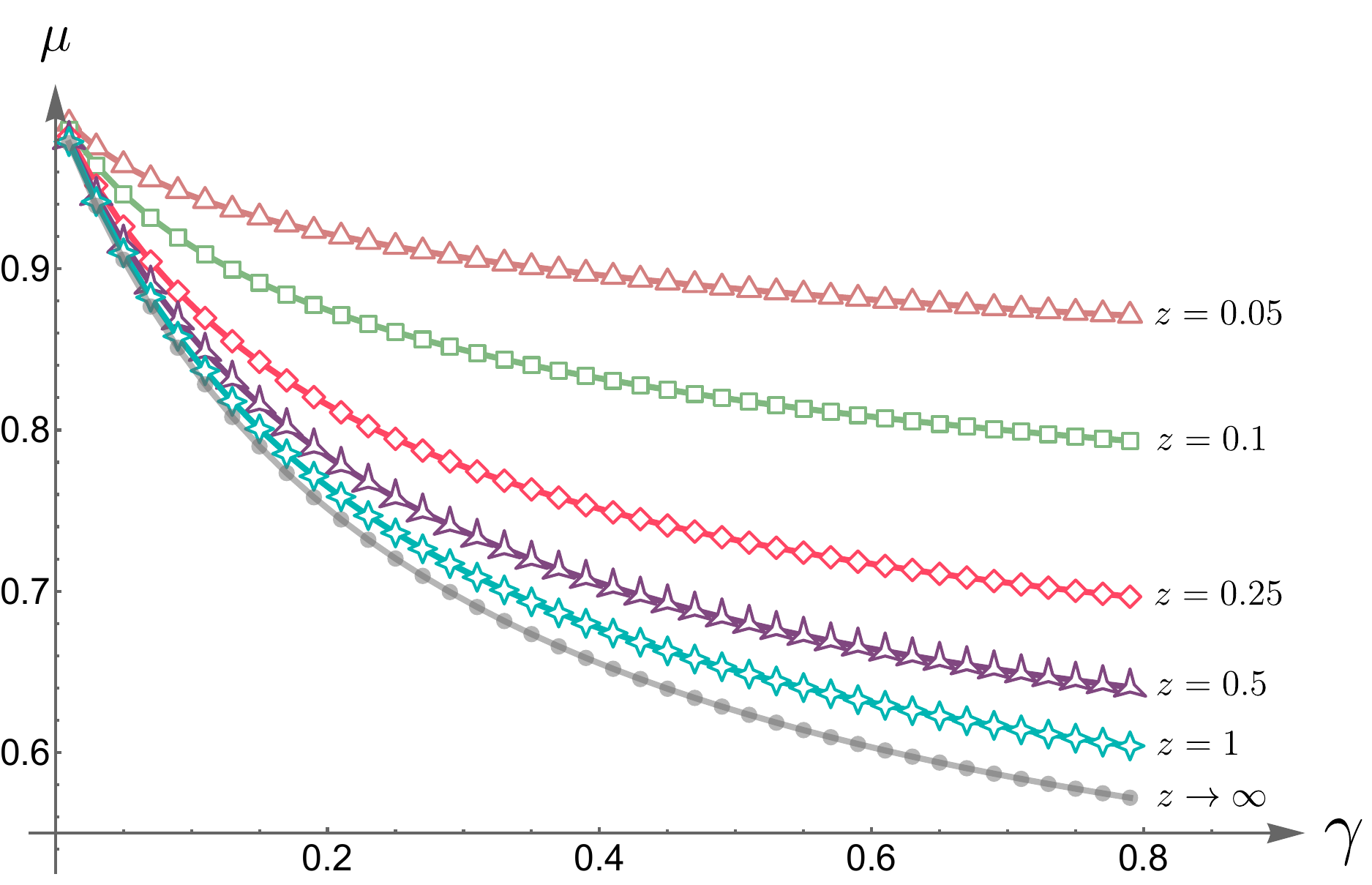}
         \caption{Variation of purity $\mu$ with the damping constant $\gamma$ to the interface for selected values of the distance $z$.}
         \label{Fi:mugammaPC}
    \end{subfigure}
    \hfill
    \begin{subfigure}[b]{0.45\textwidth}
         \centering
         \includegraphics[width=0.95\linewidth]{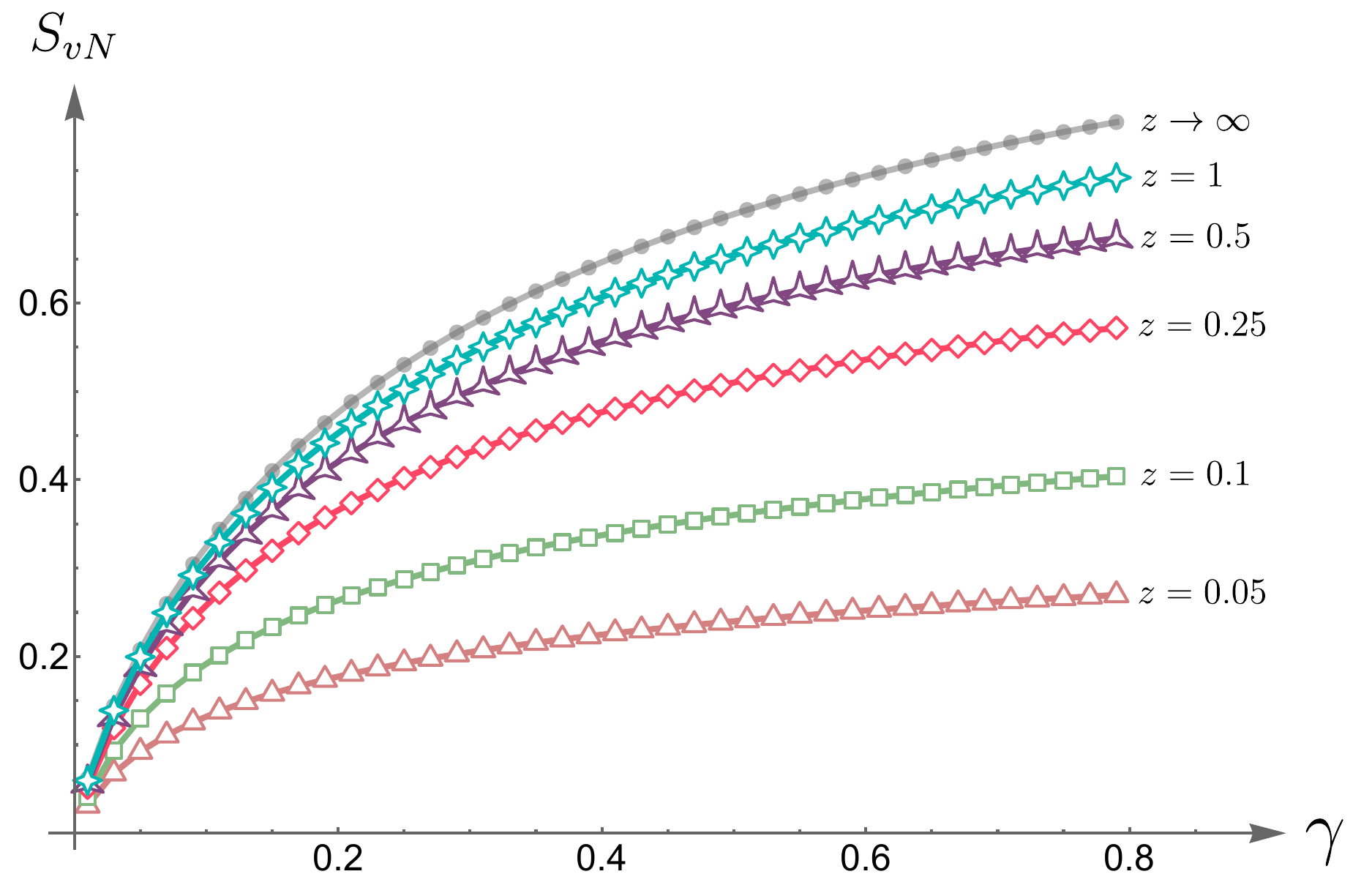}
         \caption{Variation of the von Neumann entropy $S_{vN}$ with the damping constant $\gamma$ to the interface for selected values of the distance $z$.}
         \label{Fi:svngammaPC}
    \end{subfigure}
    \hfill
    \caption{The medium is a perfect conductor. Here the distance $z$ is normalized relative to the wavelength associated with the electronic transition frequency $\omega_{\textsc{a}}$, and here $\gamma$ is $\gamma'_{\textsc{a}}$ normalized with respective to $\omega_{\textsc{a}}$. The case $z\to\infty$ corresponds to the configuration in the absence of the medium.}
    \label{Fi:sq2}
\end{figure}
In Fig.~\ref{Fi:sq2}, we  can see that the trend of decreasing effective coupling by plotting an imaginary vertical line for any value of $\gamma$. With closer distance to the boundary, the mixing is less significant. The boundary effect diminishes with larger values of $z$. The curve corresponding to $z\to\infty$ is the result obtained in unbounded space, i.e., in the absence of the conductor.

\subsubsection{Dielectric half-space}

When the conductor is replaced by the dielectric, the configuration becomes much more complicated. For one, the amplitude of each field mode outside the medium does not take on the same definite value on the boundary as in the conductor case. The field can also transmit the interface, entering the dielectric, in addition to being reflected from the interface. At the same time, the field  inside the dielectric exhibits more complicated behavior. Not all transmitted modes propagate in the vacuum; lower frequency modes tend to attenuate, so that their amplitudes exponentially decrease into vacuum at mode-dependent rates.

\begin{figure}
    \centering
    \hfill
    \begin{subfigure}[b]{0.45\textwidth}
         \centering
         \includegraphics[width=0.95\linewidth]{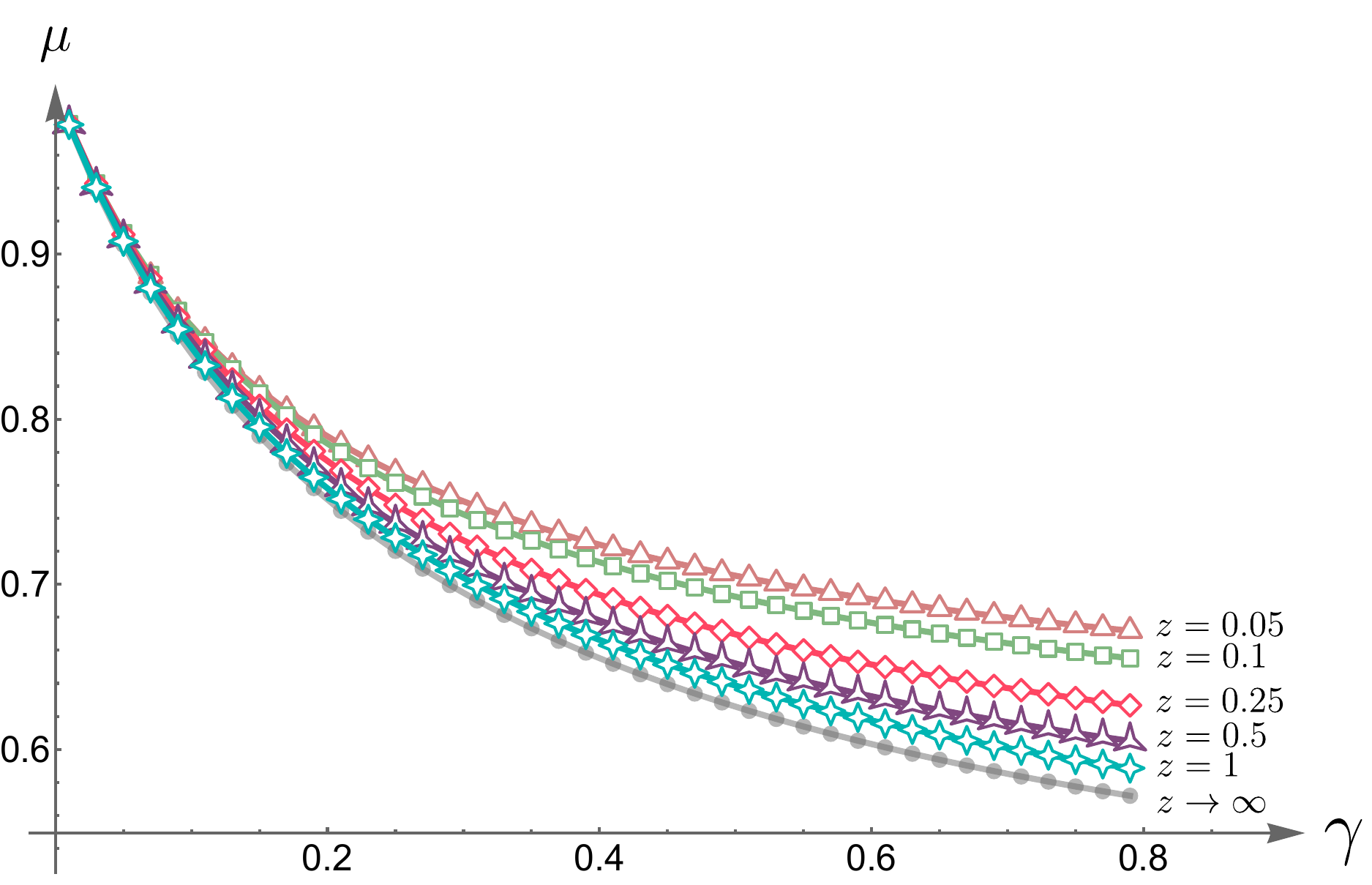}
         \caption{Variation of purity $\mu$ with the damping constant $\gamma$ to the interface for selected values of the distance $z$.}
         \label{Fi:mugammaDi}
    \end{subfigure}
    \hfill
    \begin{subfigure}[b]{0.45\textwidth}
         \centering
         \includegraphics[width=0.95\linewidth]{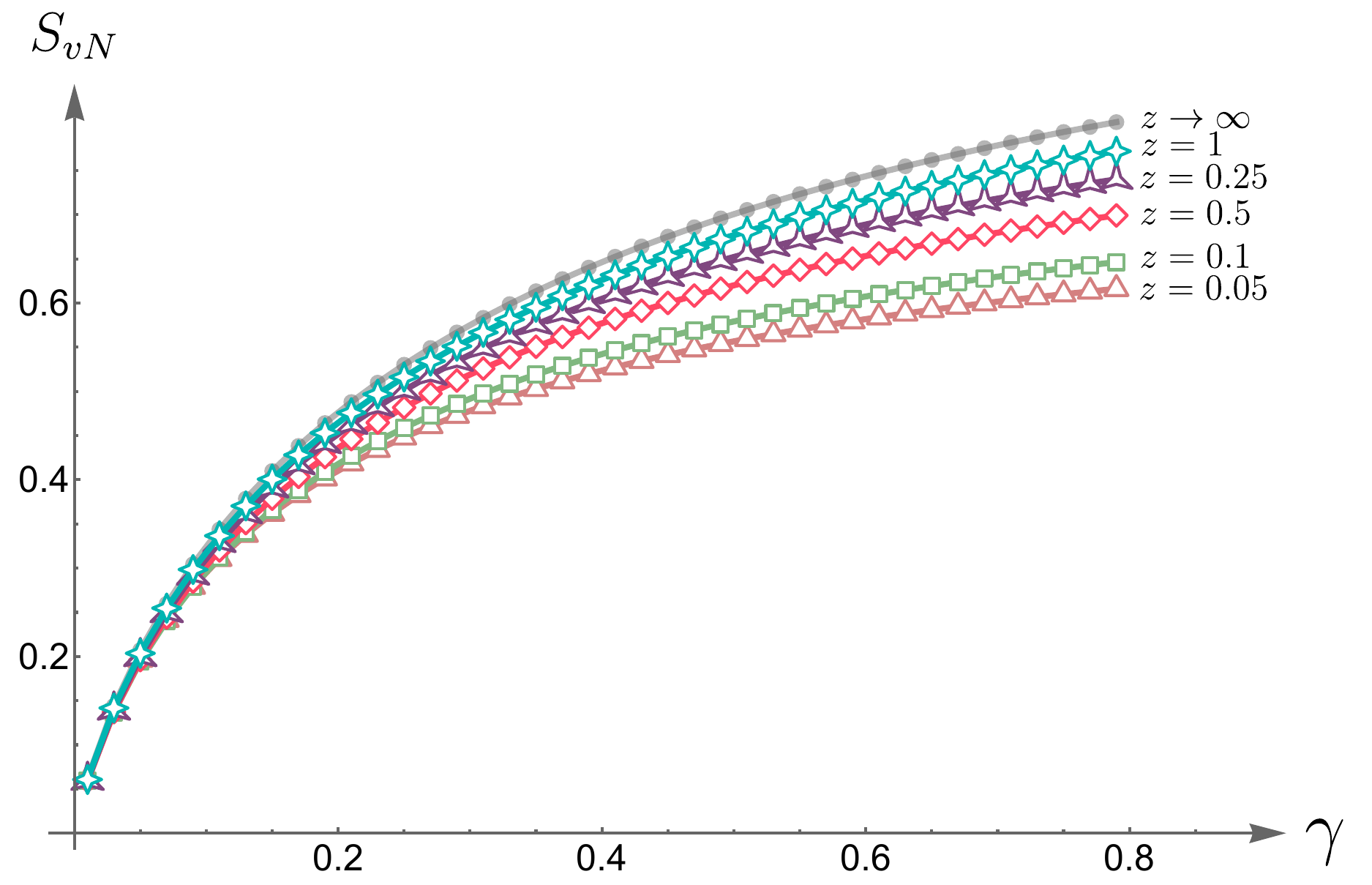}
         \caption{Variation of von Neumann entropy $S_{vN}$ with the damping constant $\gamma$ to the interface for selected values of the distance $z$.}
         \label{Fi:svngammaDi}
    \end{subfigure}
    \hfill
    \caption{The medium is a dielectric described by the single-species Lorentz model. Here the distance $z$ is normalized relative to the wavelength associated with the electronic transition frequency $\omega_{\textsc{a}}$, and  $\gamma$ is $\gamma'_{\textsc{a}}$ normalized with respective to $\omega_{\textsc{a}}$. The case $z\to\infty$ corresponds to the configuration in the absence of the medium.}
    \label{Fi:sq3}
\end{figure}

Even though in Fig.~\ref{Fi:sq3}, the curves show the same general trends as those for the conductor case in Fig.~\ref{Fi:sq2}, the curves are bunched much tighter. This is understood as follows: Unlike in the conductor case where the amplitude field or its gradient in the direction parallel to the interface is close to zero due to the boundary condition, when the medium is a dielectric, the field is continuous across the interface, taking on possible values allowed by the boundary conditions. Hence, there is no significant suppression of the effective damping/coupling constant.

\begin{figure}
    \centering
    \hfill
    \begin{subfigure}[b]{0.45\textwidth}
         \centering
         \includegraphics[width=0.95\linewidth]{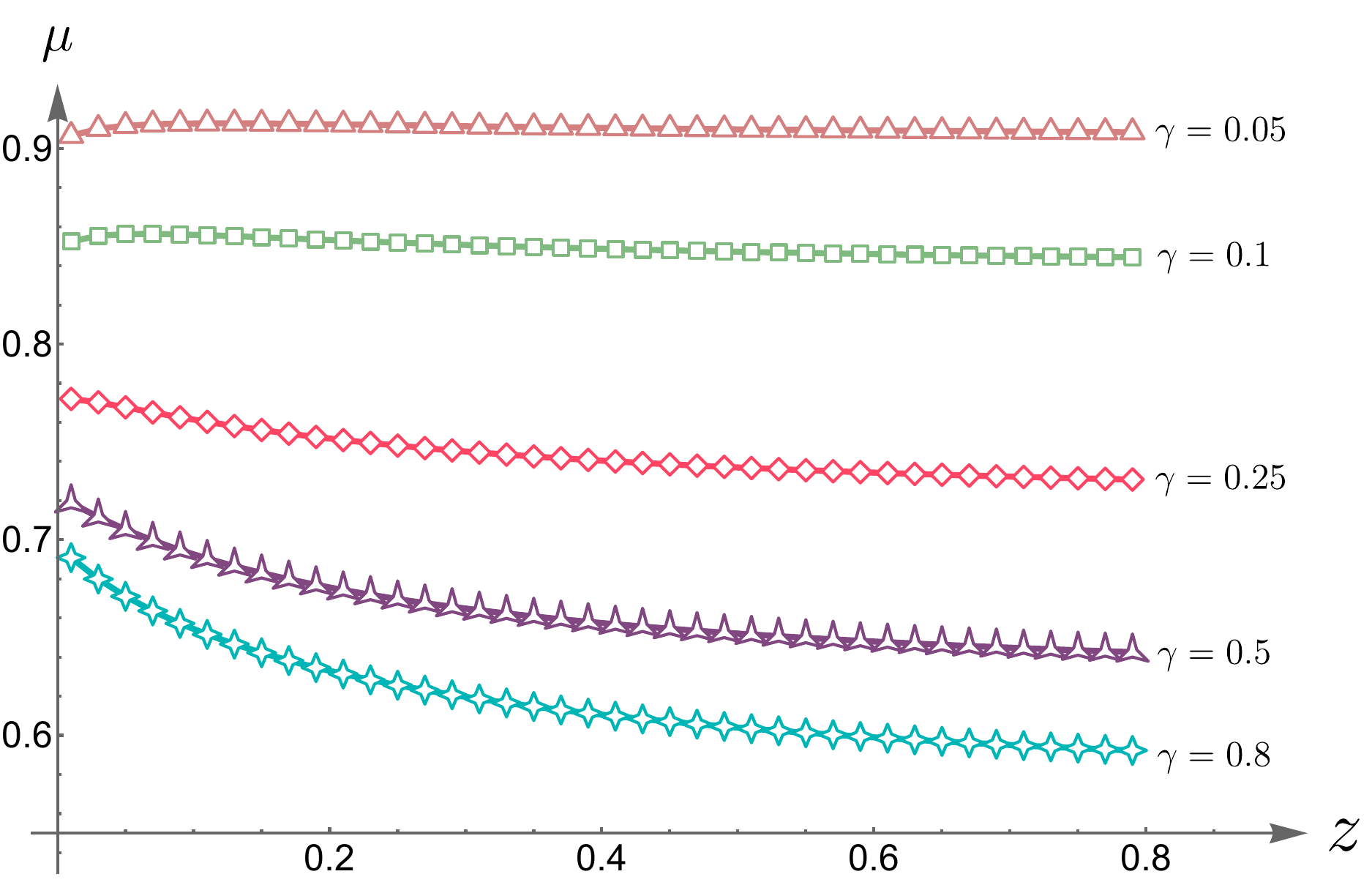}
         \caption{Variation of purity $\mu$ with the distance $z$ to the interface for selected values of the damping constant $\gamma$.}
         \label{Fi:muzDi}
    \end{subfigure}
    \hfill
    \begin{subfigure}[b]{0.45\textwidth}
         \centering
         \includegraphics[width=0.95\linewidth]{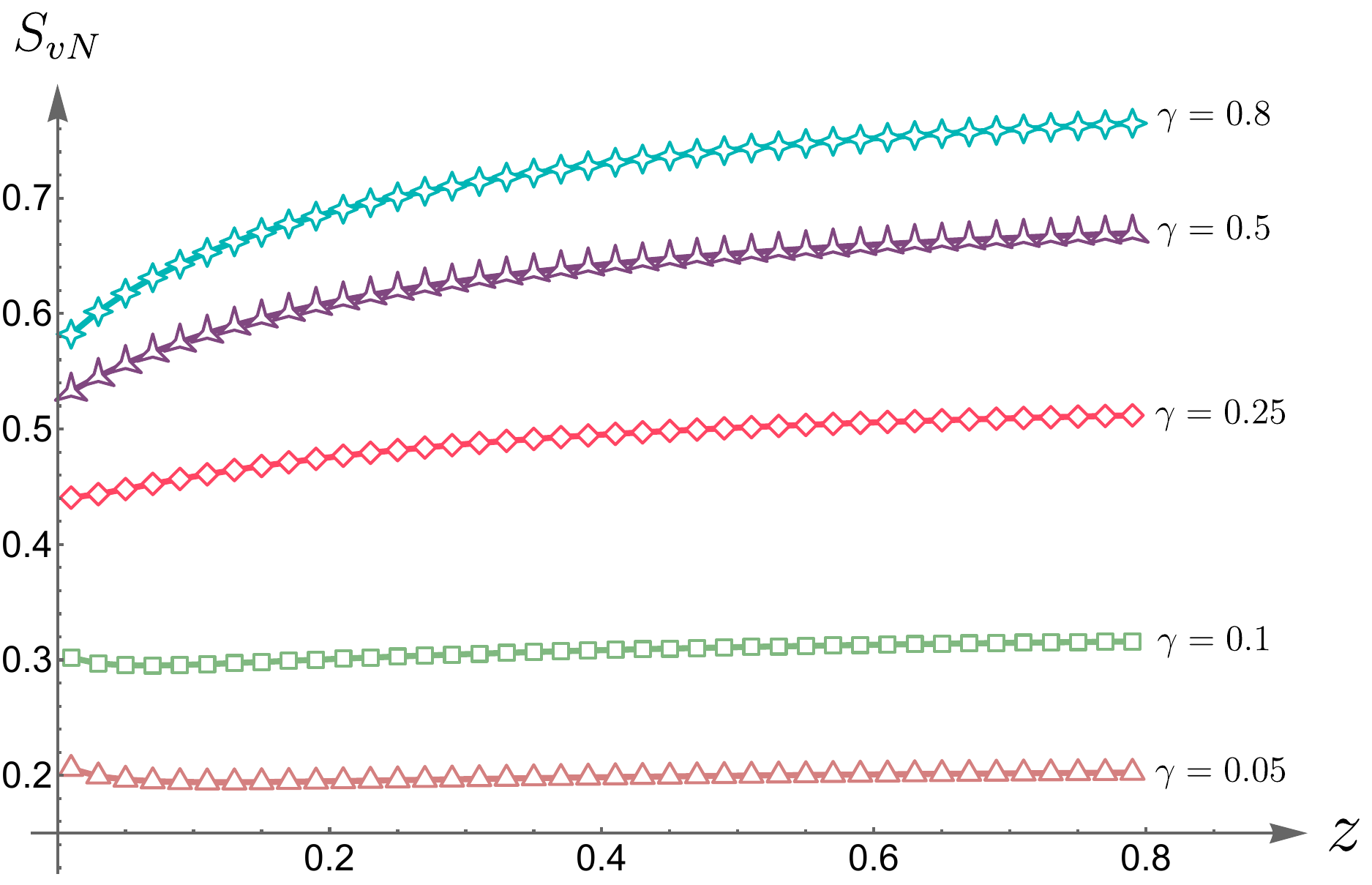}
         \caption{Variation of the von Neumann entropy $S_{vN}$ with the distance $z$ to the interface for selected values of the damping constant $\gamma$.}
         \label{Fi:svnzDi}
    \end{subfigure}
    \hfill
    \caption{The medium is a dielectric described by the single-species Lorentz model. Here the distance $z$ is normalized relative to the wavelength associated with the electronic transition frequency $\omega_{\textsc{a}}$, and $\gamma$ is $\gamma'_{\textsc{a}}$ normalized with respective to $\omega_{\textsc{a}}$. The case $z\to\infty$ corresponds to the configuration in the absence of the medium.}
    \label{Fi:sq4}
\end{figure}

\subsubsection{Behavior close to the dielectric surface}

At large values of $z$  the $z$ dependence of the entanglement measure for a  dielectric medium has essentially the same behavior as for a perfect conductor,  because far away from the boundary its effects are weak. However, at small values of $z$,  features different from the perfect conductor case are noticeable.  Foremost, the curves for either the purity $\mu$ or the von Neumann entropy does not vary monotonically. The value of $z$ where non-monotonicity in these functions shows up is approximately at $z\sim0.05$ (normalized by the wavelength $\omega_{\textsc{a}}^{-1}$) for very weak coupling $\gamma=0.05$ (normalized by $\omega_{\textsc{a}}$, i.e., $\gamma=\gamma'_{\textsc{a}}/\omega_{\textsc{a}}$).  With increasing $\gamma$ this value gradually approaches toward $z=0$ . Thus, the transition is barely visible for $\gamma=0.25$ onward.

A) Retardation?  Such non-monotonic behavior cannot be caused by the transition of the retardation effect because 1) it occurs roughly at the scale $z\sim1$ where the amount of time for field propagation becomes noticeable, and 2) usually the transition of the retardation effect  modifies the rate of change of the quantities with the distance to the interface.

B) Lennard-Jones (LJ) potential? The LJ potential is a popular model for intermolecular effects. It shows that at large distances, the intermolecular force for non-polar molecule is an attractive van der Waals force, but with shortening intermolecular separation, the force becomes repulsive, due to the Coulomb repulsive interaction between the inner electronic cores of molecules or the nuclei of atoms. The LJ  potential exhibits non-monotonic behavior with mutual separation of molecules. Now, can the non-monotonic variation in Fig.~\ref{Fi:sq4} be related to the LJ potential between the dielectric atom and the neutral atom outside the dielectric? We believe this is not the case for our model because our model does not get into the detailed atomic structures where such complex interactions arise.  Besides, this behavior seems to have been reported also in a recent paper~\cite{IP04}, so it is unlikely a numerical or physical artifact.

Rather, we conjecture that the non-monotonic behavior in these results is related to the inherently large (renormalized/smeared) quantum field fluctuations close to the boundary~\cite{SF02, SF05, NP12, BB15, SPR13, FS98, HL99}. This is one of the consequences of the uncertainty relation governing quantum systems, where, when one of the canonical variables has a well-defined value, then the uncertainty of its conjugate variable is prone to diverge. Spatial smearing, e.g. due to a fluctuating boundary, can soften the divergence to finite but large values. Since, as stressed before, at late times the  dynamics of the idf of the system atom is governed by the ambient field fluctuations, it is expected that larger field fluctuations tend to induce larger fluctuations in the atom's internal dynamics at late times. Owing to such increasingly large ambient field fluctuations the energy in the idf of the atom is anticipated to grow in the proximity of the interface.

\begin{figure}
    \centering
    \hfill
    \begin{subfigure}[b]{0.45\textwidth}
         \centering
         \includegraphics[width=0.95\linewidth]{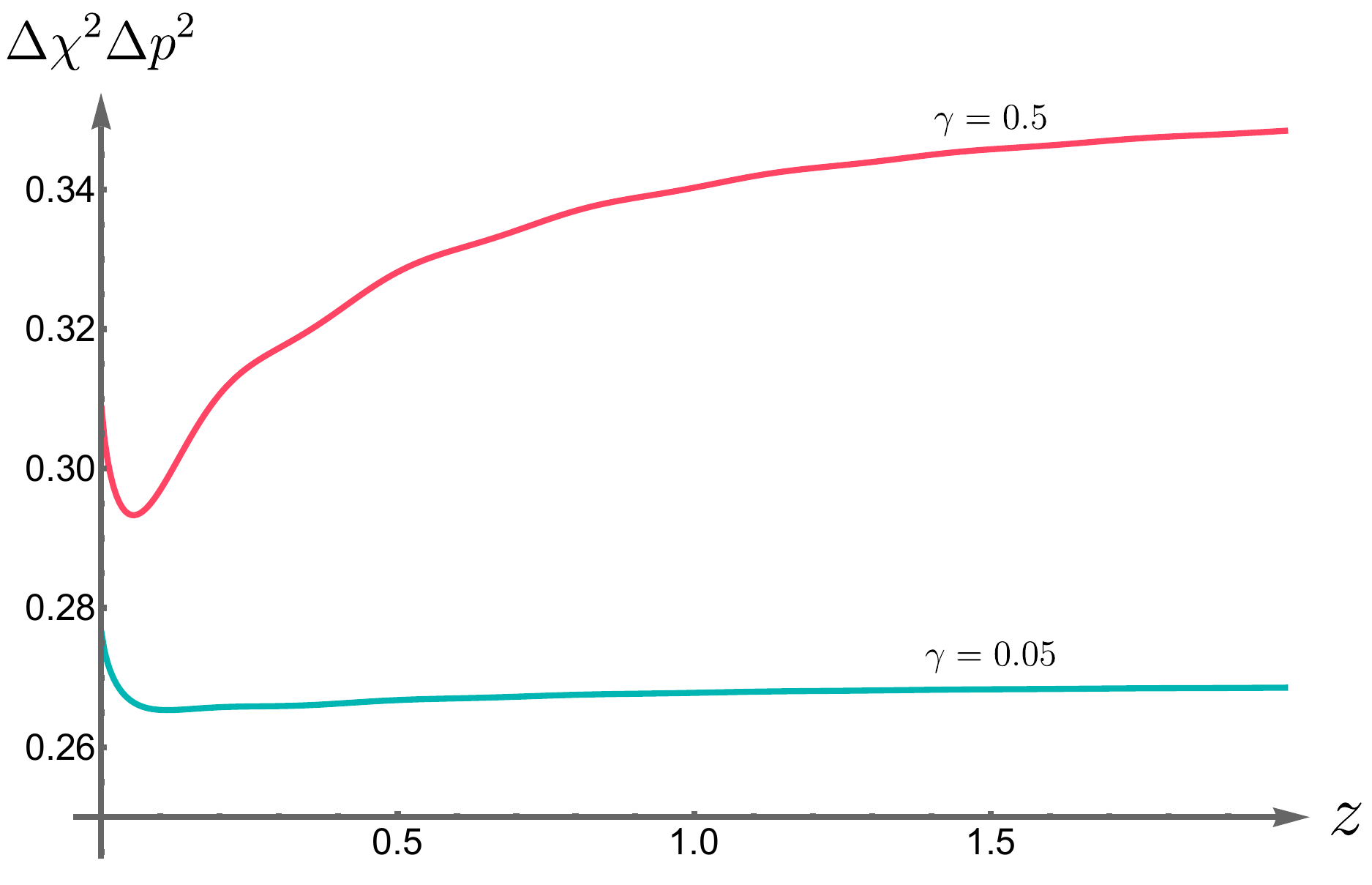}
         \caption{Variation of the Robertson-Schr\"odinger uncertainty function $\Delta\chi^2\,\Delta p^2$ with the distance $z$ to the interface for two different damping constants $\gamma$.}
         \label{Fi:SRzDi}
    \end{subfigure}
    \hfill
    \begin{subfigure}[b]{0.45\textwidth}
         \centering
         \includegraphics[width=0.95\linewidth]{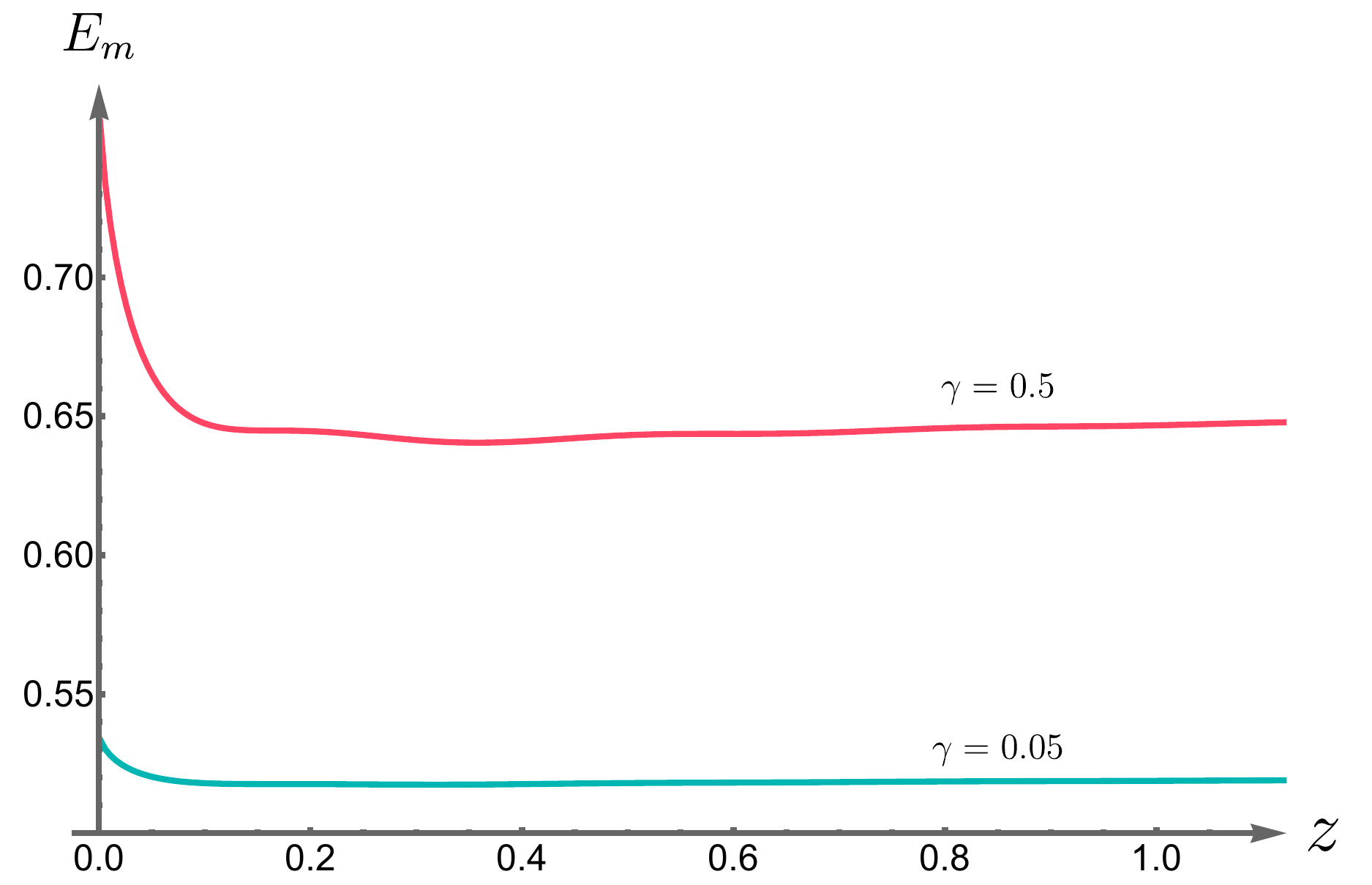}
         \caption{Variation of energy $E_{m}$ of the atom's internal degrees of freedom with the distance $z$ to the interface for two different damping constants $\gamma$.}
         \label{Fi:EmzDi}
    \end{subfigure}
    \hfill
    \caption{The dielectric medium is described by the single-species Lorentz model. Here the distance $z$ is normalized relative to the wavelength associated with the electronic transition frequency $\omega_{\textsc{a}}$, and  $\gamma$ is $\gamma'_{\textsc{a}}$ normalized with respective to $\omega_{\textsc{a}}$. The case $z\to\infty$ corresponds to the case when the medium is absent.}
    \label{Fi:sq5}
\end{figure}

To show that this interpretation makes sense, we made a plot of the mechanical energy of the atom's idf modeled as a harmonic oscillator. As seen in  Fig.~\ref{Fi:EmzDi} it increases with smaller $z$.  To ensure the self-consistency of our numerical calculations, we also calculated the Robertson-Schr\"odinger uncertainty function governing the atom's idf.  As shown in Fig.~\ref{Fi:SRzDi} it also manifests the signature non-monotonic behavior, while its values never drop below  $1/4$.

After making sure that this nonmonotonic behavior is not a numerical artifact we explore the causes of this interesting feature with some suggestive thoughts as follows.

\begin{figure}
    \centering
    \hfill
    \begin{subfigure}[b]{0.45\textwidth}
         \centering
         \includegraphics[width=0.95\linewidth]{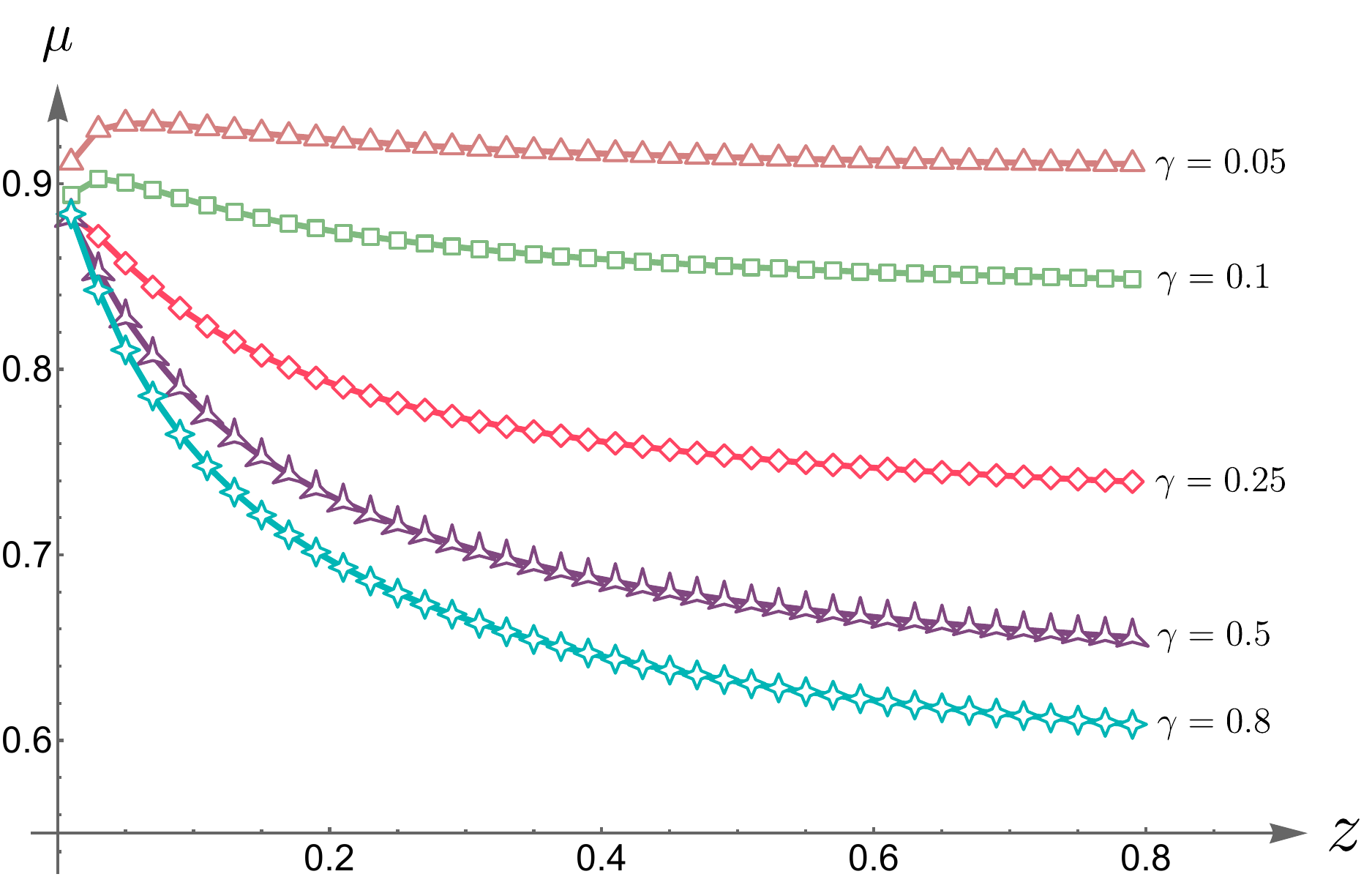}
         \caption{Variation of purity $\mu$ with the distance $z$ to the interface for selected values of the damping constant $\gamma$.}
         \label{Fi:muzDiBign}
    \end{subfigure}
    \hfill
    \begin{subfigure}[b]{0.45\textwidth}
         \centering
         \includegraphics[width=0.95\linewidth]{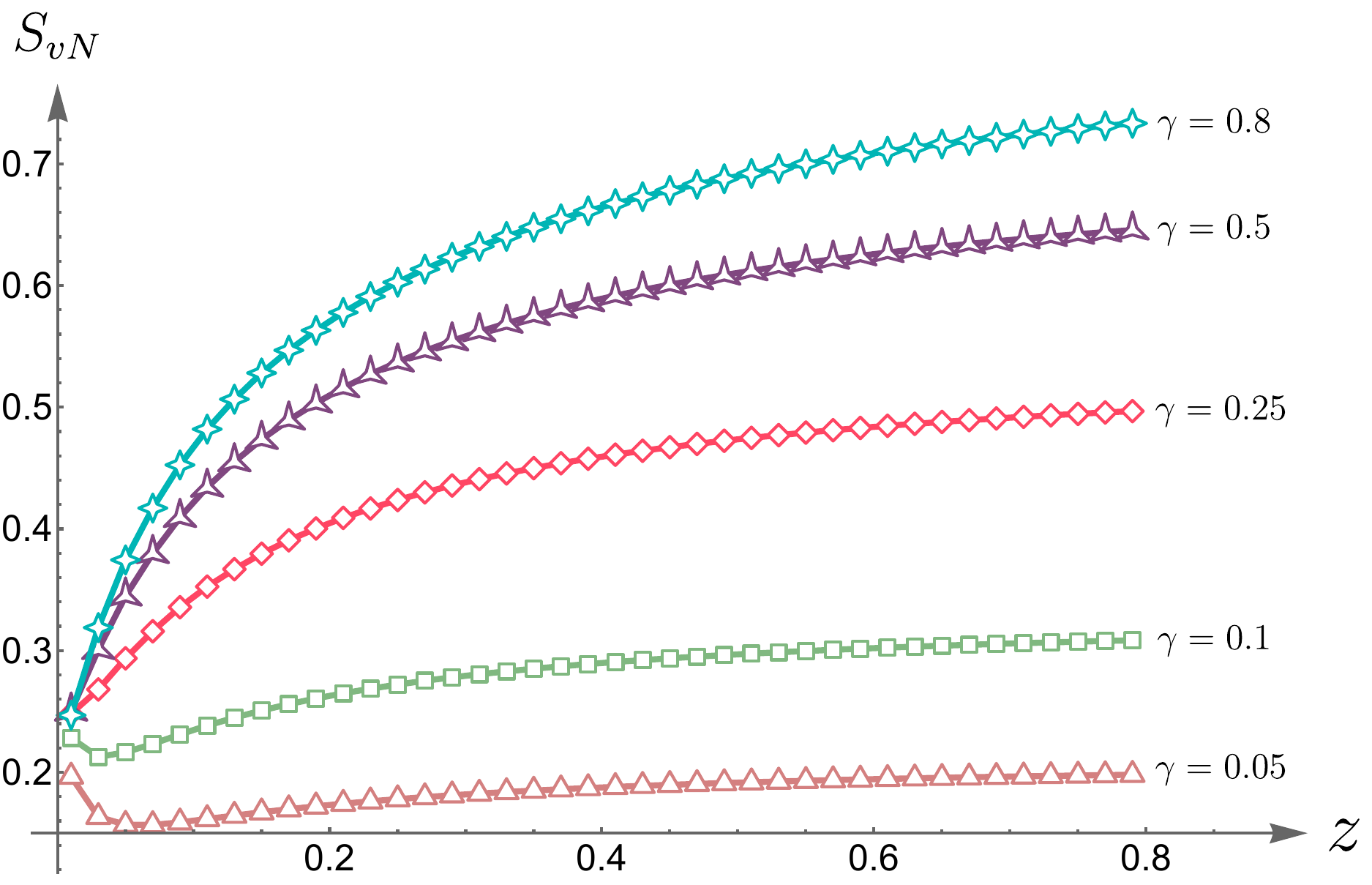}
         \caption{Variation of von Neumann entropy $S_{vN}$ with the distance $z$ to the interface for selected values of the damping constant $\gamma$.}
         \label{Fi:svnzDiBign}
    \end{subfigure}
    \hfill
    \caption{The medium is a dielectric described by the single-species Lorentz model with ten times as large the plasma frequency as used in Fig.~\ref{Fi:sq4}. Here the distance $z$ is normalized relative to the wavelength associated with the electronic transition frequency $\omega_{\textsc{a}}$, and  $\gamma$ is $\gamma'_{\textsc{a}}$ normalized with respective to $\omega_{\textsc{a}}$. The case $z\to\infty$ corresponds to the configuration in the absence of the medium.}
    \label{Fi:sq6}
\end{figure}

Another distinct feature in the dielectric case as shown in Fig.~\ref{Fi:sq4} is that at very small $z$, the curves do not converge to roughly the same value, in contrast to what is shown in Fig.~\ref{Fi:sq} for the perfect conductor case. This is also related to the boundary condition on the interface. We can see the link between these two cases by studying an intermediary case with the same set of curves as in Fig.~\ref{Fi:sq4} except that the susceptibility now increases by 10 times, so that the dielectric behaves more like a perfect conductor. In other words, the boundary condition of the field in this case better resembles that of the field in the proximity of the perfect conductor. We can see that the curves in Fig.~\ref{Fi:sq6} come together better at small $z$, i.e., in the neighborhood of the boundary, than the curves in Fig.~\ref{Fi:sq4}, but not as tight as those in Fig.~\ref{Fi:sq}. Moreover, observe that with a  larger  susceptibility   the $z$ value for the purity or  the von Neumann entropy  to make a non-monotonic transition gets closer to the interface. This  also explains why no such behavior shows when the medium is a perfect conductor. 

One relevant minor detail is that the non-monotonic behavior for fixed $\gamma$ tends to vary more prominently, when the value of susceptibility is increased. The above analysis reflects on the unnaturalness of an idealized `perfect' conductor, especially in the region very close to the boundary: Physically the renormalized/smeared field fluctuate at large values whereas mathematically, by fiat, imposition of a perfect conductor boundary condition  forces the amplitude of the field to vanish on the boundary. This is what makes the Green's function of the field vary frenziedly in the proximity of the boundary of a perfect conductor.

\section{Summary and Remarks}

{In terms of methodology, we have dwelled on the relative advantages and challenges of the functional method versus the operator method, the former by way of influence actions was treated in Paper I, the latter by way of quantum Langevin equations is used in this paper for computing the covariance matrices.   In terms of our results for atom-dielectric entanglement via an ambient quantum field from our calculation of the purity function or the von Neumann entropy, a feature which stands out is its non-monotonic behavior when the atom is placed very close to the dielectric-vacuum interface. 
We explained that this behavior is unlikely to be related to a transition to retardation effect or to the Lennard-Jones potential for the intermolecular forces. We conjecture it is a consequence of renormalized quantum field fluctuations growing unbounded at a fixed boundary. We show that similar non-monotonic feature manifests in the Robertson-Schr\"odinger uncertainty function, thus offering a grounded explanation of the physical origin of this behavior, namely, ruled by the uncertainty principle. }

In treating atom-field interactions, if the internal degrees of freedom of the atom is coupled to the field via the `derivative coupling' type, such as for the familiar case of a charge interacting with an electromagnetic field, the equation of motion for the charge sector, though local in time, but if the charge is assumed to be pointlike, contains a term of third-order time derivative. The dynamics of such a system is in general  acausal and unstable.  Such an equation is notorious for its runaway behavior, thus lacking a well-defined asymptotically equilibrium state.  This phenomenon manifests as well for a non-relativistic point electric charge coupled to the electric field~\cite{Ja98,Ro90,Ya06,Sp04,HH22}.

%If we model the internal degree of freedom of the system atom as a point dipole coupled to the derivative of the ambient field, then it is inevitable that the corresponding equation of motion, 

From a broader context of non-Markovian open systems, various pathological behavior for the equation of motion having third-order time derivatives can be circumvented if the system of interest has a minimal, nonvanishing extension~\cite{Ya06,HH22}. This minimal size is shown to be of the same order as the classical radius of the charge. Once the size of the system shrinks below this critical value, the aforementioned pathological behavior emerges. However, a non-Markovian equation of motion is very difficult to deal with, no matter analytically or numerically.  Thus, for most of the time, a local equation of motion is preferred even though it may have some undesired features. The technique of order reduction is proposed to rewrite the third-order time derivative in the equation of motion into a first-order one. The order-reduced equation of motion then has an expected relaxation behavior, leading to physically meaningful late-time dynamics.

However, from the viewpoint of the open systems, only the dissipative backreaction is applied to. The accompanying fluctuation backaction is untreated which leads to inconsistencies. In this work we proposed a  more complete order reduction scheme for  Gaussian open systems. The details for a consistent order reduction scheme are contained in Appendix B.  

Here, for the sake of full disclosure and to offer a suggestion for future investigations,  we want to bring up a somewhat disconcerting feature in the order-reduced Langevin equation Eq.~\eqref{E:feireer}. Recall the derivation of Eq.~\eqref{E:eruskjhfs}. The damping results from the system's response to the radiation field. The nonlocal expression on the left is essentially  the Li\'enard-Wiechert potential. After order reduction, the counterpart of this potential has a different form, so it raises the question of whether it alters the radiation pattern. A similar concern applies to the fluctuation backaction. The right hand side of Eq.~\eqref{E:feireer} does not look like a Lorentz force as seen on the right hand side of Eq.~\eqref{E:eruskjhfs}, casting doubts on whether the order-reduced equation is a sufficiently accurate description of the charged system coupled the electromagnetic field or its alike? We conjecture that a fully order-reduced scheme might offer an adequate description in the limit of weak atom-field interaction, where the internal dynamics of the atom possesses a constricted but prominent Breit-Wigner peak about the transition frequency. As addressed earlier, a non-Markovian description~\cite{HH22} of such a system is preferable because it is inherently stable as long as the system has a finite extension greater than a critical scale on the order of the classical charge radius. Thus, it would be interesting to compare its predictions with those of the order-reduced description. We hope to address this issue in a future paper.

\quad\\
\noindent{\bf Acknowledgments} 
J.-T. Hsiang is supported by the National Science and Technology Council of Taiwan, R.O.C. under Grant No.~NSTC 113-2112-M-011-001-MY3. He thanks Professor Larry Ford for helpful discussions on the phenomena of large quantum fluctuations near a surface. The present work began in 2013-2014 when both authors visited the physics department of Fudan University, Shanghai, China. They are thankful for the warm reception of Professor Y. S. Wu, then the Director of the new Fudan Center for Theoretical Physics. In 2024-2025 B.-L. Hu enjoyed the gracious hospitality of colleagues at the Institute of Physics, Academia Sinica and at the National Center for Theoretical Sciences at Tsing Hua University, Hsinchu, Taiwan, R.O.C. where this work was finished.

%\newpage
\appendix

% \newpage  \appendix

\section{Nonequilibrium Quantum Correlations in Multipartite Quantum Brownian Motion}\label{App:ebfsjdbfs}
Here we derive the nonequilibrium covariance matrix for multipartite quantum Brownian motion in its final state, as has been shown in~\cite{FH11}. Suppose that $N$ identical quantum harmonic oscillators of mass $M$ and frequency $\Omega$ couple to a common bath $B$. The dynamics of this quantum Brownian system follows the quantum Langevin equation,
\begin{equation}\label{E:ouree}
	M\,\ddot{\bm{X}}(t)+M\Omega^{2}\,\bm{X}(t)-g^{2}\int_{0}^{t}\!dt'\;\bm{G}_{R}^{(B)}(t-t')\cdot\bm{X}(t')=g\,\bm{\xi}(t)\,,
\end{equation}
driven by the quantum noise $\bm{\xi}$, in which the displacement vector $\bm{X}(t)$ of the oscillators is an $N$ dimensional column vector. The retarded kernel $\bm{G}_{R}^{(B)}$ of the bath is an $N\times N$ symmetric matrix. The coupling strength between the Brownian system and the bath is denoted by $g$.

The formal solution of \eqref{E:ouree} can be found with the help of the Laplace transformation of the Langevin equation \eqref{E:ouree},
\begin{align}
	\bm{X}(t)=\bm{X}_{h}(t)+\frac{g}{M}\int_{0}^{t}\!ds\;\bm{G}_{R}^{(X)}(t-s)\cdot\bm{\xi}(s)\,,
\end{align}
where $\bm{X}_{h}(t)$ is the homogeneous solution to the Langevin equation and are determined by the initial conditions. It usually decays with time due to the presence of damping, which is the consequence of the interaction between the Brownian system and the bath. The Fourier transform of the retarded function $\bm{G}_{R}^{(X)}$ of the Brownian system is defined by
\begin{equation}
	\widetilde{\bm{G}}_{R}^{(X)}(\omega)=\Bigl[M\bigl(\Omega^{2}-\omega^{2}\bigr)\,\bm{I}-g^{2}\,\widetilde{\bm{G}}_{R}^{(B)}(\omega)\Bigr]^{-1}\,.
\end{equation}
From this we can construct the elements of the covariance matrix of the Brownian system. For example, the elements $\bm{V}_{XX}(t)=\dfrac{1}{2}\langle\bigl\{\bm{X}(t),\,\bm{X}^{T}(t)\bigr\}\rangle$ are given by
\begin{align}\label{E:djksf}
	\bm{V}_{XX}(t)&=\cdots+\frac{g^{2}}{M^{2}}\int_{0}^{t}\!ds\,ds'\;\bm{G}_{R}^{(X)}(t-s)\cdot\dfrac{1}{2}\langle\bigl\{\bm{\xi}(s),\,\bm{\xi}^{T}(s')\bigr\}\rangle\cdot\bm{G}_{R}^{(X)}{}^{T}(t-s')\,,
\end{align}
where $\cdots$ represents contributions related to $\bm{X}_{h}$ and are usually exponentially small at late times so we will ignore them. The superscript $T$ means transpose. If the bath is Gaussian, then the expectation value of the quantum noise $\bm{\xi}$ is related to the noise kernel of the bath, $\dfrac{1}{2}\langle\bigl\{\bm{\xi}(s),\,\bm{\xi}^{T}(s')\bigr\}\rangle=\bm{G}_{H}^{(B)}(s,s')$. We note that due to the presence of the retarded kernel $\bm{G}_{R}^{(X)}$, we can set the upper limit of the integrals in \eqref{E:djksf} to $\infty$, and \eqref{E:djksf} becomes
\begin{align}
	\bm{V}_{XX}(t)&=\frac{g^{2}}{M^{2}}\int_{-\infty}^{t}\!ds\,ds'\;\bm{G}_{R}^{(X)}(s)\cdot\bm{G}_{H}^{(B)}(s-s')\cdot\bm{G}_{R}^{(X)}{}^{T}(s')\,,
\end{align}
where we have made the change of the variable. In general, $\bm{G}_{H}^{(B)}(s,s')$ is not always time-translationally invariant, but it has be shown that in the limit $t\to\infty$ it becomes time translationally invariant. Thus in the limit $t\to\infty$ we have
\begin{align}
	\bm{V}_{XX}(\infty)&=\frac{g^{2}}{M^{2}}\int_{-\infty}^{\infty}\frac{d\omega}{2\pi}\;\widetilde{\bm{G}}_{R}^{(X)*}(\omega)\cdot\widetilde{\bm{G}}_{H}^{(B)}(\omega)\cdot\widetilde{\bm{G}}_{R}^{(X)}{}(\omega)\label{E:uererss}\\
	&=\frac{g^{2}}{M^{2}}\int_{-\infty}^{\infty}\frac{d\omega}{2\pi}\;\coth\frac{\beta\omega}{2}\;\widetilde{\bm{G}}_{R}^{(X)*}(\omega)\cdot\operatorname{Im}\widetilde{\bm{G}}_{R}^{(B)}(\omega)\cdot\widetilde{\bm{G}}_{R}^{(X)}{}(\omega)\,,\label{E:bjss}
\end{align}
in which $\widetilde{\bm{G}}_{R}^{(X)*}(\omega)=\widetilde{\bm{G}}_{R}^{(X)}(-\omega)$ because $\bm{G}_{R}^{(X)}(s)$ is real, and the initial state of the bath has been assumed thermal. The latter implies a fluctuation-dissipation relation for the bath
\begin{equation}
	\widetilde{\bm{G}}_{H}^{(B)}(\omega)=\coth\frac{\beta\omega}{2}\,\operatorname{Im}\widetilde{\bm{G}}_{R}^{(B)}(\omega)\,.
\end{equation}
Eq.~\eqref{E:uererss} makes the reference to both the noise kernel of the bath and the dissipation kernel of the system.

To eliminate explicit reference to the noise kernel of the bath, we will use a matrix identity 
\begin{align}
	\bm{A}^{-1}-\bm{B}^{-1}&=\bm{A}^{-1}\bigl(\bm{B}-\bm{A}\bigr)\bm{B}^{-1}\,,&&\text{for the matrices $\bm{A}$, $\bm{B}$}
\end{align}
to re-write $\widetilde{\bm{G}}_{R}^{(X)*}(\omega)\cdot\operatorname{Im}\widetilde{\bm{G}}_{R}^{(B)}(\omega)\cdot\widetilde{\bm{G}}_{R}^{(X)}{}(\omega)$ in \eqref{E:bjss} as,
\begin{align}
	\frac{g^{2}}{M}\,\widetilde{\bm{G}}_{R}^{(X)*}(\omega)\cdot\operatorname{Im}\widetilde{\bm{G}}_{R}^{(B)}(\omega)\cdot\widetilde{\bm{G}}_{R}^{(X)}{}(\omega)=\operatorname{Im}\widetilde{\bm{G}}_{R}^{(X)}(\omega)\,.
\end{align} 
Thus $\bm{V}_{XX}(\infty)$ becomes
\begin{align}\label{E:enskjns1}
	\bm{V}_{XX}(\infty)&=\frac{1}{M}\,\operatorname{Im}\int_{-\infty}^{\infty}\frac{d\omega}{2\pi}\;\coth\frac{\beta\omega}{2}\,\widetilde{\bm{G}}_{R}^{(X)}(\omega)\,.
\end{align}
Likewise we can show that, with $\bm{P}=M\dot{\bm{X}}$,
\begin{align}
	\bm{V}_{PP}(\infty)&=M^{2}\langle\dot{\bm{X}}(t)\cdot\dot{\bm{X}}^{T}(t)\rangle=M\,\operatorname{Im}\int_{-\infty}^{\infty}\frac{d\omega}{2\pi}\;\omega^{2}\coth\frac{\beta\omega}{2}\,\widetilde{\bm{G}}_{R}^{(X)}(\omega)\,,\label{E:enskjns2}\\
	\bm{V}_{XP}(\infty)&=\frac{1}{2}\,\langle\{\bm{X}(t),\,\bm{P}^{T}(t)\}\rangle=0\,,\label{E:enskjns3}
\end{align}
because the integrand in $\bm{V}_{XP}$ is odd in $\omega$. From the derivation, we see that the stationary nature of bath's initial state definitively bestows simplicity of eqs.~\eqref{E:enskjns1}, \eqref{E:enskjns2} and \eqref{E:enskjns3}.

\section{Consistent order reduction scheme in open quantum systems} \label{App:bebeirie}

In this appendix which is a standalone module by itself, we present an order reduction scheme applicable to open quantum systems.  Order reduction on an equation of motion which contains third or higher order derivatives in time is a familiar subject in treating radiation reaction problems, such as in the Abraham-Lorentz-Dirac equation in electromagnetic radiation theory.  In Sec II.5 we have pointed out that there are inconsistencies in the traditional order reduction scheme, namely, that it is only implemented on the dissipative backaction on the left hand side of \eqref{E:bieura}, but not on the fluctuation backaction on its right hand side.   For simplicity, we leave the $\cdots$ terms in \eqref{E:bieura} untouched for the moment.  The second issue relating to the renormalization of the parameters is subtler. To arrive at Eq.~\eqref{E:oeuhd}, we have performed a mass renormalization to rewrite $M\,\ddot{x}$ in \eqref{E:bieura} as $M_{\textsc{a}}\ddot{x}$. In so doing, we also need to adjust the other parameters in \eqref{E:oeuhd} accordingly. However, after order reduction, Eq.~\eqref{E:oeuhd} is transformed into Eq.~\eqref{E:bdjuer}, the derivation of which, as is well known in treating Brownian motion in the open system framework, uses frequency renormalization. We need to examine whether a particular order reduction scheme retains the consistency of the renormalizations of these two parameters.

To address these issues in the context of open systems, it is instructive to briefly review the derivation of the equation of motion for Brownian motion, which has the same form as \eqref{E:bdjuer}. Let us start from
\begin{equation}\label{E:oritfhg}
    M_{\textsc{a}}\,\ddot{\hat{\chi}}^{(n)}(t)+M_{\textsc{a}}\Omega'^2\,\hat{\chi}^{(n)}(t)-g'^2_{\textsc{f}}\int^t\!\!dt'\;G_{\mathrm{R,0}}^{(\phi_h)}(\bm{z}_n,t;\bm{z}_{n},t')\,\hat{\chi}^{(n')}(t')+\cdots=g'_{\textsc{f}}\,\hat{\phi}_h(\bm{z}_n,t)\,,
\end{equation}
which is similar to \eqref{E:bktrr}. Here, it would be more illustrative to choose a different coupling constant $g'_{\textsc{f}}$ and a different frequency $\Omega'$ from those used in \eqref{E:bieura}. Plugging Eq.~\eqref{E:itrrrt} into \eqref{E:oritfhg}, and implementing an integration by parts, we find
\begin{equation}
    M_{\textsc{a}}\,\ddot{\hat{\chi}}^{(n)}(t)+M_{\textsc{a}}\Omega'^2\,\hat{\chi}^{(n)}(t)-\frac{g'^2_{\textsc{f}}}{2\pi}\,\delta(0)\,\hat{\chi}^{(n)}(t)+\frac{g'^2_{\textsc{f}}}{4\pi}\,\dot{\hat{\chi}}^{(n)}(t)+\cdots=g'_{\textsc{f}}\,\hat{\phi}_h(\bm{z}_n,t)\,,
\end{equation}
for $t>0$. If we define
\begin{align}\label{E:uuerw}
    M_{\textsc{a}}\,\delta\Omega^2&=-\frac{g'^2_{\textsc{f}}}{2\pi}\,\delta(0)\,,&&\text{such that}&\Omega'^2_{\textsc{a}}&=\Omega'^2+\delta\Omega^2\,,&2M_{\textsc{a}}\gamma'_{\textsc{a}}&=\frac{g'^2_{\textsc{f}}}{4\pi}\,,
\end{align}   
respectively as the frequency renormalization and damping constant, we obtain an expression similar to  Eq.~\eqref{E:bdjuer}
\begin{equation}\label{E:iufere}
    M_{\textsc{a}}^{\vphantom{2}}\,\ddot{\hat{\chi}}^{(n)}(t)+M_{\textsc{a}}^{\vphantom{2}}\Omega'^2_{\textsc{a}}\,\hat{\chi}^{(n)}(t)+2M_{\textsc{a}}^{\vphantom{2}}\gamma'_{\textsc{a}}\,\dot{\hat{\chi}}^{(n)}(t)+\cdots=g_{\textsc{f}}\,\hat{\phi}_h(\bm{z}_n,t)\,.
\end{equation} 
The salient point here is that the mass parameter $M_{\textsc{a}}$ in Eq.~\eqref{E:oritfhg} does not need renormalization; only the frequency parameter gets renormalized for the system described by Eq.~\eqref{E:oritfhg}.

In contrast, Eq.~\eqref{E:oeuhd} is obtained from
\begin{equation}\label{E:dkbe}
    M^{\vphantom{2}}\,\ddot{\hat{\chi}}^{(n)}(t)+M\Omega^2\,\hat{\chi}^{(n)}(t)+\frac{g^2_{\textsc{f}}}{2\pi}\,\delta(0)\,\ddot{\hat{\chi}}^{(n)}(t)-\frac{g^2_{\textsc{f}}}{4\pi}\,\dddot{\hat{\chi}}^{(n)}(t)+\cdots=g_{\textsc{f}}\,\dot{\hat{\phi}}_h(\bm{z}_n,t)\,.
\end{equation}
for $t>0$. Eq.~\eqref{E:dkbe} in turn is derived from Eq.~\eqref{E:bieura} after we carry out two integrations by parts, and then assign
\begin{align}\label{E:iitier}
    \delta M&=\frac{g^2_{\textsc{f}}}{2\pi}\,\delta(0)\,,&&\text{such that} &M_{\textsc{a}}&=M+\delta M\,,&M^{\vphantom{2}}_{\textsc{a}}\Omega^2_{\textsc{a}}&=M\Omega^2\,.
\end{align}
Thus, not only is the mass renormalized, the natural frequency is necessarily altered from $\Omega$ to $\Omega_{\textsc{a}}$

Now we examine whether it is possible to cast Eq.~\eqref{E:oeuhd} in a form like Eq.~\eqref{E:iufere} by order reduction, that is, $\ddot{\hat{\chi}}^{(n)}(t)=-\Omega^2_{\textsc{x}}\,\hat{\chi}^{(n)}(t)$ for some frequency parameter $\Omega_{\textsc{x}}$ to be determined later, such that the resulting expression is compatible with the open-system formulation. By this we mean whether the equation of motion, order-reduced from \eqref{E:oeuhd}, can be derived from a nonlocal equation similar to Eq.~\eqref{E:oritfhg}. Two simplest choices of $\Omega_{\textsc{x}}$ will be the unrenormalized frequency $\Omega$ and the physical frequency $\Omega_{\textsc{a}}$.

Order reduction of Eq.~\eqref{E:oeuhd} gives
\begin{equation}\label{E:eiudbgewr}
    M^{\vphantom{2}}_{\textsc{a}}\,\ddot{\hat{\chi}}^{(n)}(t)+M^{\vphantom{2}}_{\textsc{a}}\Omega^2_{\textsc{a}}\,\hat{\chi}^{(n)}(t)+\frac{g^2_{\textsc{f}}}{4\pi}\,\Omega^2_{\textsc{x}}\,\dot{\hat{\chi}}^{(n)}(t)+\cdots=g_{\textsc{f}}\,\dot{\hat{\phi}}_h(\bm{z}_n,t)\,.
\end{equation}
If the left hand side of Eq.~\eqref{E:eiudbgewr} is compatible with that of Eq.~\eqref{E:iufere}, that is, derivable from an expression like Eq.~\eqref{E:oritfhg}, then the renormalization is applied only to the frequency, while  leaving $M_{\textsc{a}}$ untouched, as outlined in Eq.~\eqref{E:uuerw}. Thus the expression $M^{\vphantom{2}}_{\textsc{a}}\Omega^2_{\textsc{a}}$ in Eq.~\eqref{E:eiudbgewr} in principle can be decomposed into
\begin{equation}\label{E:eirieirw}
    M^{\vphantom{2}}_{\textsc{a}}\Omega^2_{\textsc{a}}=M^{\vphantom{2}}_{\textsc{a}}\Omega''^2+M^{\vphantom{2}}_{\textsc{a}}\delta\Omega''^2
\end{equation}
where $\Omega''$ is some unrenormalized bare frequency. It in turn  implies that the corresponding frequency renormalization $\delta\Omega''^2$, together with
\begin{equation}
    2M_{\textsc{a}}\gamma''_{\textsc{a}}=\frac{g^2_{\textsc{f}}}{4\pi}\,\Omega^2_{\textsc{x}}\,,
\end{equation}
ought to be obtained from an expression like
\begin{equation}
    -g^2_{\textsc{f}}\Omega^2_{\textsc{x}}\int^t\!\!dt'\;G_{\mathrm{R,0}}^{(\phi_h)}(\bm{z}_n,t;\bm{z}_{n},t')\,\hat{\chi}^{(n)}(t')\,.
\end{equation} 
In addition, $\Omega''$ can be identified as $\Omega$ in Eq.~\eqref{E:bieura}.

Now, suppose the above conjecture does hold. Then we find
\begin{align}
    M^{\vphantom{2}}_{\textsc{a}}\delta\Omega''^2=-\frac{g^2_{\textsc{f}}}{2\pi}\,\Omega^2_{\textsc{x}}\,\delta(0)\,.
\end{align}
This implies that the decomposition \eqref{E:eirieirw} can be written as
\begin{equation}\label{E:bgetri}
    M\Omega^2=M^{\vphantom{2}}_{\textsc{a}}\Omega^2_{\textsc{a}}=M^{\vphantom{2}}_{\textsc{a}}\Omega''^2-\frac{g^2_{\textsc{f}}}{2\pi}\,\Omega^2_{\textsc{x}}\,\delta(0)=M^{\vphantom{2}}\Omega''^2+\frac{g^2_{\textsc{f}}}{2\pi}\,\delta(0)\,\Omega''^2-\frac{g^2_{\textsc{f}}}{2\pi}\,\delta(0)\,\Omega^2_{\textsc{x}}\,,
\end{equation}
with the help of \eqref{E:iitier}. It is clear to see that Eq.~\eqref{E:bgetri} holds identically if and only if $\Omega''=\Omega_{\textsc{x}}=\Omega$, such that Eq.~\eqref{E:eiudbgewr} can be derived from a nonlocal equation,
\begin{equation}\label{E:vgeutd}
    M^{\vphantom{2}}_{\textsc{a}}\,\ddot{\hat{\chi}}^{(n)}(t)+M^{\vphantom{2}}_{\textsc{a}}\Omega^2\,\hat{\chi}^{(n)}(t)-g^2_{\textsc{f}}\Omega^2\int^t\!\!dt'\;G_{\mathrm{R,0}}^{(\phi_h)}(\bm{z}_n,t;\bm{z}_{n},t')\,\hat{\chi}^{(n)}(t')+\cdots=g_{\textsc{f}}\,\dot{\hat{\phi}}_h(\bm{z}_n,t)\,.
\end{equation}
This is essentially of the same form as Eq.~\eqref{E:oritfhg}. Comparing Eq.~\eqref{E:vgeutd} with Eq.~\eqref{E:oritfhg}, we are prompted to make such assignments as
\begin{align}
    \Omega'&=\Omega\,,&&\text{and}&g'_{\textsc{f}}&=g_{\textsc{f}}\,\Omega\,,
\end{align}
that the left hand side of Eq.~\eqref{E:eiudbgewr} is cast as
\begin{equation}
    M^{\vphantom{2}}_{\textsc{a}}\,\ddot{\hat{\chi}}^{(n)}(t)+M^{\vphantom{2}}_{\textsc{a}}\Omega^2_{\textsc{a}}\,\hat{\chi}^{(n)}(t)+M^{\vphantom{2}}_{\textsc{a}}\Omega^2_{\textsc{a}}\,\tau^{\vphantom{2}}_{\textsc{a}}\,\dot{\hat{\chi}}^{(n)}(t)+\cdots=g_{\textsc{f}}\,\dot{\hat{\phi}}_h(\bm{z}_n,t)\,,
\end{equation}    
with
\begin{equation}
    M^{\vphantom{2}}_{\textsc{a}}\Omega^2_{\textsc{a}}\,\tau^{\vphantom{2}}_{\textsc{a}}=\frac{g^2_{\textsc{f}}}{4\pi}\,\Omega^2\,,
\end{equation}
and it in turn  requires $\Omega^2_{\textsc{a}}\,\tau^{\vphantom{2}}_{\textsc{a}}=\Omega^2\,\tau$ according to the definition of $\tau$.

As we have seen so far, the dynamics described by Eq.~\eqref{E:bieura} introduces mass renormalization or effective mass. Such a change inevitably adjusts the other parameters such as $\Omega$
\begin{align}
    M^{\vphantom{2}}_{\textsc{a}}\Omega^2_{\textsc{a}}&=M\Omega^2\,,
\end{align}   
in the equation. In this context, if the prescription of order reduction on the Lifshitz equation is to be made meaningful in the open systems framework, then one is naturally lead to $g'_{\textsc{f}}=g_{\textsc{f}}\,\Omega$. This seems to provide a different perspective, with implication that Eq.~\eqref{E:oeuhd} might not be  so physically meaningful, depending on how one interprets the parameter renormalization, because $\tau$ or $g_{\textsc{f}}$ in the equation takes on the original, ``unrenormalized'' value,
\begin{equation}
    \frac{g'^2_{\textsc{f}}}{4\pi}=2M^{\vphantom{2}}_{\textsc{a}}\gamma'_{\textsc{a}}=M^{\vphantom{2}}_{\textsc{a}}\Omega^2_{\textsc{a}}\,\tau^{\vphantom{2}}_{\textsc{a}}=\frac{g^2_{\textsc{f}}}{4\pi}\,\Omega^2\,.
\end{equation}
It may be more preferable to use $\tau_{\textsc{a}}$ or $g'_{\textsc{f}}$ in the equation of motion.

Next, we ask:  what if we use $\Omega_{\textsc{a}}$ in the reduction of order, instead of $\Omega$? Following oour previous derivations, we will come to an equation of the form 
\begin{equation}
     M^{\vphantom{2}}_{\textsc{a}}\,\ddot{\hat{\chi}}^{(n)}(t)+M^{\vphantom{2}}\Omega^2_{\textsc{a}}\,\hat{\chi}^{(n)}(t)-g^2_{\textsc{f}}\Omega^2_{\textsc{a}}\int^t\!\!dt'\;G_{\mathrm{R,0}}^{(\phi_h)}(\bm{z}_n,t;\bm{z}_{n},t')\,\hat{\chi}^{(n)}(t')+\cdots=g_{\textsc{f}}\,\dot{\hat{\phi}}_h(\bm{z}_n,t)\,.
\end{equation}  
This is not the typical expression we have for the Brownian motion because different masses are involved. Thus, this choice $\Omega_{\textsc{a}}$ is undesirable.

So far, we have worked out the procedures to apply order reduction to the left hand side of Eq.~\eqref{E:bieura} in the context of open systems, 
\begin{align}
    &&&M\,\ddot{\hat{\chi}}^{(n)}(t)+M\Omega^2\,\hat{\chi}^{(n)}(t)+g_{\textsc{f}}^2\int^t\!\!dt'\;\partial_tG_{\mathrm{R,0}}^{(\phi_h)}(\bm{z}_n,t;\bm{z}_{n},t')\,\dot{\hat{\chi}}^{(n)}(t')+\cdots\label{E:eoitoet}\\
    &\Rightarrow&&M^{\vphantom{2}}_{\textsc{a}}\,\ddot{\hat{\chi}}^{(n)}(t)+M^{\vphantom{2}}_{\textsc{a}}\Omega^2\,\hat{\chi}^{(n)}(t)-g^2_{\textsc{f}}\Omega^2\int^t\!\!dt'\;G_{\mathrm{R,0}}^{(\phi_h)}(\bm{z}_n,t;\bm{z}_{n},t')\,\hat{\chi}^{(n)}(t')+\cdots\,.\label{E:euyeir}
\end{align}
with $g'_{\textsc{f}}=g_{\textsc{f}}\,\Omega$. As mentioned earlier, for the sake of consistency, order reduction should be implemented on the right hand side as well. A hint can be seen if we write Eq.~\eqref{E:eoitoet} as
\begin{align}\label{E:swreew}
    &\hphantom{=}M\,\ddot{\hat{\chi}}^{(n)}(t)+M\Omega^2\,\hat{\chi}^{(n)}(t)+g_{\textsc{f}}^2\int^t\!\!dt'\;\partial_tG_{\mathrm{R,0}}^{(\phi_h)}(\bm{z}_n,t;\bm{z}_{n},t')\,\dot{\hat{\chi}}^{(n)}(t')+\cdots\notag\\
    &=M\,\ddot{\hat{\chi}}^{(n)}(t)+M\Omega^2\,\hat{\chi}^{(n)}(t)+g_{\textsc{f}}^2\int^t\!\!dt'\;\partial_t^2G_{\mathrm{R,0}}^{(\phi_h)}(\bm{z}_n,t;\bm{z}_{n},t')\,\hat{\chi}^{(n)}(t')+\cdots
\end{align}
by integration by parts and then changing the derivative $\partial_{t'}$ to $\partial_{t}$ in the integrand. The Fourier transform of $\partial_t^2G_{\mathrm{R,0}}^{(\phi_h)}(\bm{z}_n,t;\bm{z}_{n},t')$ is $-\omega^2\tilde{G}_{\mathrm{R,0}}^{(\phi_h)}(\bm{z}_n,\bm{z}_{n};\omega)$. Comparing with the corresponding expression in \eqref{E:euyeir}, as far as backactions are concerned, we may identify the time derivative in \eqref{E:swreew} with $\Omega$ to find an expression equivalent to that from order reduction. It then follows that we may cast as $g_{\textsc{f}}\Omega\,\hat{\phi}_h(\bm{z}_n,t)$ the expression $g_{\textsc{f}}\,\dot{\hat{\phi}}_h(\bm{z}_n,t)$ on the right hand side of \eqref{E:vgeutd}, which is a contribution associated with the fluctuation backaction. Therefore,  a proposed consistent treatment of order reduction on   Gaussian open systems is first to put the left hand side of Eq.~\eqref{E:eruskjhfs} into a ``canonical form" like Eq.~\eqref{E:swreew} and then to replace the time derivative on $\bm{G}_{\mathrm{R}}^{(\phi_h)}$ and $\bm{\xi}$ by $\Omega$.

In so doing, when  order reduction is consistently applied to both the dissipative backaction and the fluctuation backaction,  we arrive at a nonlocal equation of the form
\begin{equation}\label{E:whewes1}
    M^{\vphantom{2}}_{\textsc{a}}\,\ddot{\hat{\chi}}^{(n)}(t)+M^{\vphantom{2}}_{\textsc{a}}\Omega^2\,\hat{\chi}^{(n)}(t)-g^2_{\textsc{f}}\Omega^2\int^t\!\!dt'\;G_{\mathrm{R,0}}^{(\phi_h)}(\bm{z}_n,t;\bm{z}_{n},t')\,\hat{\chi}^{(n')}(t')+\cdots=g_{\textsc{f}}\Omega\,\hat{\phi}_h(\bm{z}_n,t)\,,
\end{equation}
or
\begin{equation}\label{E:whewes2}
    M^{\vphantom{2}}_{\textsc{a}}\,\ddot{\hat{\chi}}^{(n)}(t)+M^{\vphantom{2}}_{\textsc{a}}\Omega^2_{\textsc{a}}\,\hat{\chi}^{(n)}(t)+M^{\vphantom{2}}_{\textsc{a}}\Omega^2_{\textsc{a}}\,\tau^{\vphantom{2}}_{\textsc{a}}\,\dot{\hat{\chi}}^{(n)}(t)+\cdots=g_{\textsc{f}}\Omega\,\hat{\phi}_h(\bm{z}_n,t)\,,
\end{equation}   
with $g'_{\textsc{f}}=g_{\textsc{f}}\,\Omega$. Here $M_{\textsc{a}}$, $\Omega_{\textsc{a}}$, $\tau_{\textsc{a}}$, and $g'_{\textsc{f}}$ are physical parameters.

To get a better idea of the self-consistency of this prescription, let us work out the quantum relaxation dynamics of Eq.~\eqref{E:whewes1} or \eqref{E:whewes2}. Since they have the form of the equation of motion for typical Brownian motion, the reduced system described by either of them has an asymptotic equilibrium state as $t\to\infty$. For simplicity, we assume that the dielectric is not present and only one neutral atom is placed at $\bm{z}$ in the ambient field $\phi$, which is initially prepared in its thermal state at temperature $\beta^{-1}$ in unbounded space. As such, we do not need to worry about the $\cdots$ term in \eqref{E:whewes1} or \eqref{E:whewes2}. The Hadamard function of the internal degree of freedom $\chi$ at large $t$, $t'$ is then given by~\cite{}
\begin{align}\label{E:eouhre}
    \frac{1}{2}\,\langle\bigl\{\hat{\chi}(t),\,\hat{\chi}(t')\bigr\}\rangle&=G_{\mathrm{H}}^{(\chi_h)}(t,t')\notag\\
    &=\cdots+g_{\textsc{f}}^2\Omega^2\int_0^t\!ds\!\int^{t'}_0\!ds'\;G_{\mathrm{R}}^{(\chi_h)}(t-s)\,G_{\mathrm{R}}^{(\chi_h)}(t-s)\,G_{\mathrm{H},0}^{(\phi_h)}(\bm{z},s;\bm{z},s')
\end{align}    
where $G_{\mathrm{H},0}^{(\phi_h)}$ are the Hadamard function of $\phi_h$, in this case satisfying the source-free, wave equation of free field, $(\partial_t^2-\bm{\nabla}^2)\phi_h=0$. Thus by construction we already have
\begin{equation}\label{E:bdkgier}
    \tilde{G}_{\mathrm{H},0}^{(\phi_h)}(\bm{z},\bm{z};\omega)=\coth\frac{\beta\omega}{2}\,\operatorname{Im}\tilde{G}_{\mathrm{R}}^{(\phi_h)}(\bm{z},\bm{z};\omega)\,.
\end{equation}
The retarded function $G_{\mathrm{R}}^{(\chi_h)}(t-s)$ obeys \eqref{E:whewes1} or \eqref{E:whewes2} with their right hand sides replaced by a delta function $\delta(t-s)$. The $\cdots$ expression in Eq.~\eqref{E:eouhre} is the contribution that depends on the initial condition and it is exponentially small at large $t$, $t'$. Following the arguments in~\cite{}, we find that in the limits $t$, $t'\to\infty$
\begin{align}\label{E:fkbetu}
    G_{\mathrm{H}}^{(\chi_h)}(t,t')=g_{\textsc{f}}^2\Omega^2\int_{-\infty}^{\infty}\!\frac{d\omega}{2\pi}\;\lvert\tilde{G}_{\mathrm{R}}^{(\chi_h)}(\omega)\rvert^2\,\tilde{G}_{\mathrm{H},0}^{(\phi_h)}(\bm{z},\bm{z};\omega)\,e^{-i\omega(t-t')}\,,
\end{align}
where $\tilde{G}_{\mathrm{R}}^{(\chi_h)}(\omega)$ in this case is given by
\begin{equation}
    \tilde{G}_{\mathrm{R}}^{(\chi_h)}(\omega)=\Bigl\{M_{\textsc{a}}\bigl[-\omega^2+\Omega^2-\frac{g^2_{\textsc{f}}\Omega^2}{M_{\textsc{a}}}\,\tilde{G}_{\mathrm{R},0}^{(\phi_h)}(\bm{z},\bm{z};\omega)\bigr]\Bigr\}^{-1}\,.
\end{equation}
After some algebraic manipulations we have
\begin{equation}
    \operatorname{Im}\tilde{G}_{\mathrm{R}}^{(\chi_h)}(\omega)=\lvert\tilde{G}_{\mathrm{R}}^{(\chi_h)}(\omega)\rvert^2\,g^2_{\textsc{f}}\Omega^2\,\operatorname{Im}\tilde{G}_{\mathrm{R},0}^{(\phi_h)}(\bm{z},\bm{z};\omega)\,.
\end{equation}
Thus Eq.~\eqref{E:fkbetu} implies
\begin{align}\label{E:soieuro}
    \tilde{G}_{\mathrm{H}}^{(\chi_h)}(\omega)=g_{\textsc{f}}^2\Omega^2\lvert\tilde{G}_{\mathrm{R}}^{(\chi_h)}(\omega)\rvert^2\,\tilde{G}_{\mathrm{H},0}^{(\phi_h)}(\bm{z},\bm{z};\omega)=\coth\frac{\beta\omega}{2}\,\operatorname{Im}\tilde{G}_{\mathrm{R}}^{(\chi_h)}(\omega)\,.
\end{align}
This is the fluctuation-dissipation relation (FDR) for the internal degree of freedom of the system atom described by the order-reduced equation of motion. The relation has the same form as the conventional FDRs in the equilibrium or close to equilibrium setting. In addition, it has a form identical to the one for the field $\phi$ in Eq.~\eqref{E:bdkgier}. A major difference of Eq.~\eqref{E:soieuro} from the latter ones is that Eq.~\eqref{E:soieuro} is obtained after dynamical equilibration of the internal dynamics of the atom. This example shows that if we had not consistently applied the order reduction to both the fluctuation and dissipation backactions in the open-system framework, we may not obtain a simple and familiar nonequilibrium FDR~\eqref{E:soieuro} for the internal degree of freedom of the atom, because the factor $g_{\textsc{f}}^2\Omega^2$ would not be canceled. In fact, this is exactly what will happen. If the order reduction is applied to only the dissipative backreaction, then the resulting FDR has an additional factor that will depend on the coupling constant and the parameters of the reduced system which lessens the universality appeal of FDRs.

Using the order reduction prescription  proposed here, we can cast Eq.~\eqref{E:eruskjhfs} into
\begin{equation}\label{E:bekoer}
	M_{\textsc{a}}\ddot{\bm{X}}(t)+M_{\textsc{a}}\Omega^{2}\,\bm{X}(t)-g'^{2}_{\textsc{f}}\int_{0}^{t}dt'\;\bm{G}_{\mathrm{R}}^{(\phi_h)}(t,t')\cdot\bm{X}(t')=g'_{\textsc{f}}\,\bm{\xi}^{(\phi)}(t)\,,
\end{equation}
where $g'_{\textsc{f}}=g_{\textsc{f}}\Omega$ is the effective/physical coupling strength.

\end{document}